\documentclass[%
 reprint,
 amsmath,amssymb,
 aps,
floatfix
]{revtex4-2}

\usepackage{graphicx}
\usepackage{dcolumn}
\usepackage{bm}
\usepackage{bbm}
\usepackage{orcidlink}
\begin{document}

\title{Quantum random walks on a beam splitter array}

\author{M.I. Estrada-Delgado \orcidlink{0000-0003-3019-2878}}
\email{iestrada@tec.mx}
\affiliation{Tecnológico de Monterrey, Escuela de Ingeniería y Ciencias, Carr. Lago de Guadalupe Km. 3.5, CP. 52926, Estado de México, Mexico}
\author{Z. Blanco-Garcia \orcidlink{0000-0002-4612-7934}}%
\email{zblanco@tec.mx}
\affiliation{%
 Tecnológico de Monterrey, Escuela de Ingeniería y Ciencias, Carr. Lago de Guadalupe Km. 3.5, CP. 52926, Estado de México, Mexico}%

\date{\today}

\begin{abstract}
We present a comprehensive matrix representation of a beam splitter array, incorporating multiple input and output channels. We propose treating each beam splitter as rotational matrices of a $2n-{th}$-dimensional space. With these operators, the matrix that describes the entire square array and, consequently, the final probability distribution of an input photon state can be calculated. Furthermore, square and non-square arrays are explored using the same approach, encompassing certain interferometer arrays, the quantum quincunx, and the discrete quantum Zeno effect.

\end{abstract}

\maketitle



\section{\label{sec:level1}Introduction}

Aharonov et al. introduced the concept of quantum random walks in their 1993 paper \cite{Aha93}, in analogy to classical walks. If we consider the evolution of a quantum state in a beam splitter array, we can observe a manifestation of quantum phenomena. For this reason, it is of great interest to the quantum community.  
In recent years, investigations around this subject have increased \cite{Kem03, Per10, Sch12, Cha14, Sar19, Tra02, May11,Pat07,Ton23,Zha20}. One reason for this increase is the potential technological application in quantum computer \cite{Kem03}, cryptography, quantum information and other related fields \cite{Abd21, Cha14}. 
In 2019, Sarkar et al. implemented the quantum random walk to generate a quantum random number generator that can be used, for example, in cryptography protocols \cite{Sar19}. Others groups are performing experimental implementations of the quantum random walk algorithms \cite{Per10, Du03, Do05}. For instance, Travaglione et. al implemented the quantum random walk in an ion trap quantum computer to compare it with its classical counterparts \cite{Tra02}. 
Also, Jiangfeng Du et. al. implemented quantum random walk algorithms in a nuclear magnetic resonance quantum computer \cite{Du03}. 
The community hopes that using quantum algorithms can reduce the computing times of certain classical computer problems\cite{She03}. 

In this work, we utilize the properties of the beam splitters \cite{Yur86,Bla16,Zha20} and draw inspiration from the Elitzur and Vaidman model \cite{Eli93}, to propose a new set of operators associated with them. These operators facilitate the construction of the general beam splitter array operator. Moreover, we put forth a general expression for such operator in a certain Hilbert space that depends on the dimension of the array. Once such an operator is built, it enables the calculation of the probability evolution of a photon in several particular cases.

The paper is organized as follows:  In Sec. \ref{sec:level2}, we review the matrix beam splitter representation. 
In Sec.  \ref{sec:level3}, we introduced our model of a $2\times2$ beam splitter array and the general operator of an $n\times n$ array in certain Hilbert spaces. In Sec. \ref{sec:level4}, we apply our model to some interesting problems. Finally, the conclusion are included in Sec.  \ref{sec:level5}.

\section{\label{sec:level2} Beam Splitter}

\begin{figure}[!ht]
	\centering	\includegraphics[width=0.5\linewidth]{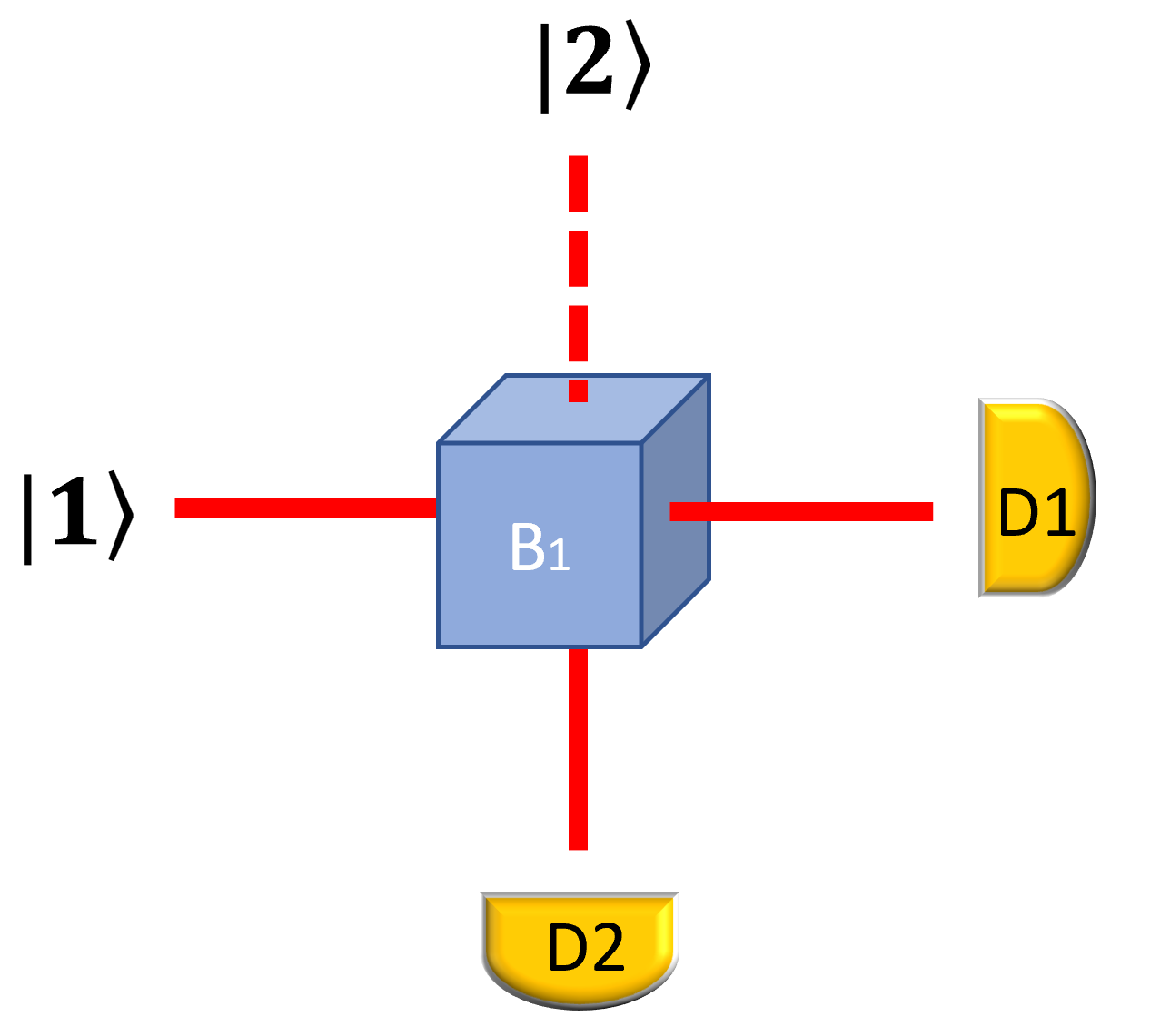}
	\caption{A general beam splitter described by the rotation matrix in \ref{eqB} is depicted. $D1$ and $D2$ represent the detectors at the two outputs of the beam splitter.}
	\label{BS1}
\end{figure}

Consider a beam splitter described by reflection and transmission coefficients $R$ and $T$, respectively. When a photon passes through this optical device, it can be sent through the horizontal arm as state $\vert 1 \rangle$ or the vertical arm as state $\vert 2 \rangle$, as shown in Figure \ref{BS1}. The matrix representations of these two states are as follows:

\begin{equation}\label{}
	\vert 1 \rangle = \begin{pmatrix}
		1 \\
		0
	\end{pmatrix}, \hskip2ex
	\vert 2 \rangle = \begin{pmatrix}
		0 \\
		1
	\end{pmatrix}.
\end{equation}

\noindent
The eigenstates $\vert 1 \rangle $ and $\vert 2 \rangle $ represent the basis states of the positional Hilbert Space.

According to \cite{Yur86} and \cite{Bla16}, these optical devices can be described by a scattering matrix operator defined as

\begin{equation}\label{eqB}
B=\begin{pmatrix}
	\cos\theta & i\sin\theta \\
	i \sin\theta & \cos\theta
\end{pmatrix}
\end{equation}

\noindent
such that, if we send an individual photon in one of the arms, it is possible to describe the photon's evolution within this device, resulting in $R:T = \sin^2\theta: \cos^2\theta$. Note that the perfect beam splitter can be recovered if $\theta = \frac{\pi}{4}$.  in the equation \ref{eqB}.

\begin{figure}[ht]
	\centering	\includegraphics[width=0.6\linewidth]{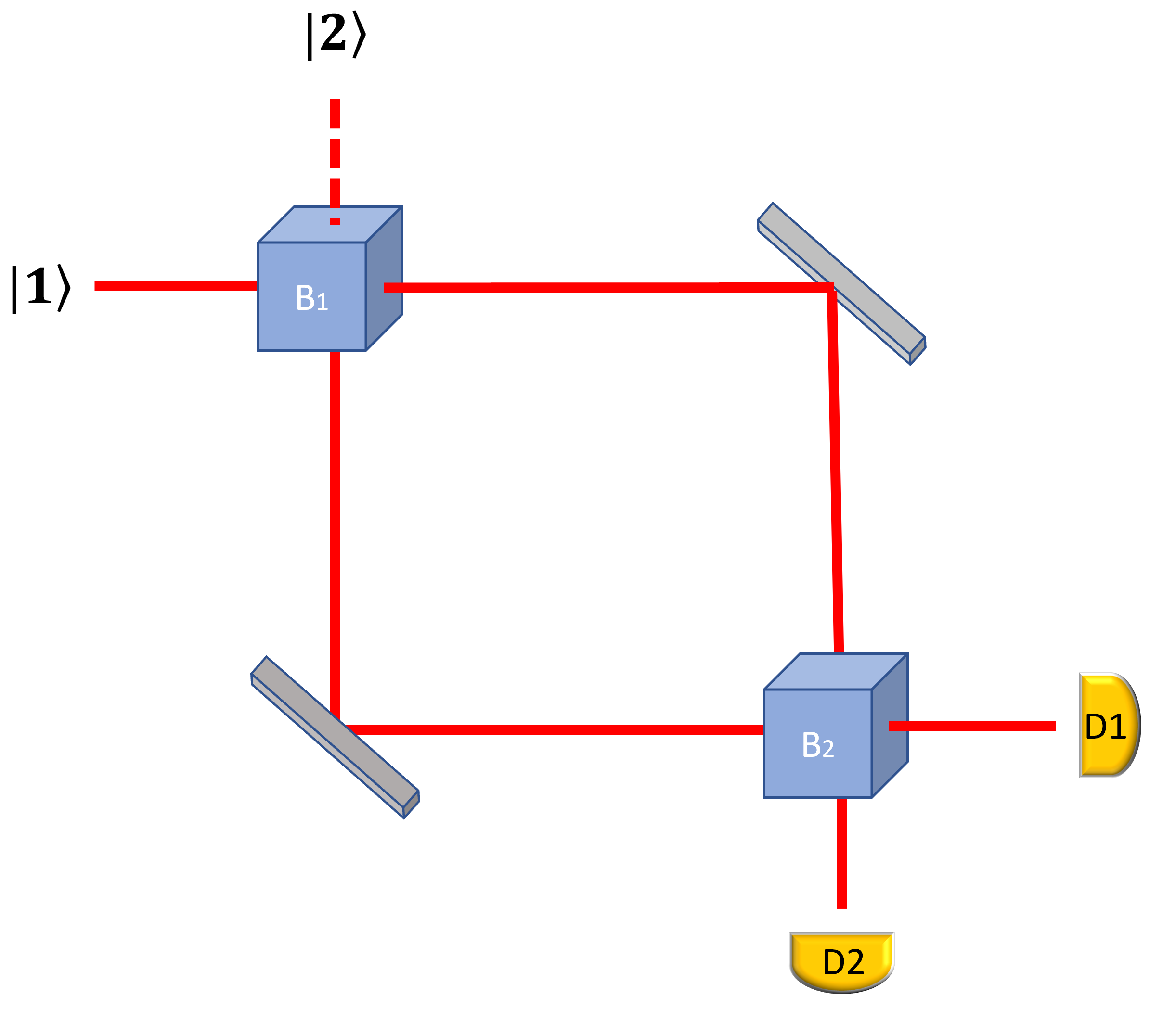}
	\caption{A Mach-Zehnder interferometer consisting of two beam splitters and two mirrors. This is a conventional model with two inputs and two outputs.}
	\label{figMZ}
\end{figure}

Using this matrix description of a beam splitter, it is possible to analyze the state evolution of a photon in 
different two-channel intereforemeters; such as a Mach-Zehnder interferometer (see Figure \ref{figMZ}). 

\noindent
The Mach-Zehnder interferometer has two beam splitters and two mirrors. The mirrors can be described using equation \ref{eqB} by considering $\theta = \frac{\pi}{2}$. When we send an individual photon through the horizontal arm with an initial state $\vert \psi_{ini} \rangle = \vert 1 \rangle$, it has a certain probability of taking either the vertical or horizontal path; the subsequent evolution performs interference of each of these arms. The final state is given as $\vert \psi_{fin} \rangle  = -2 \sin\theta \cos\theta \vert 1 \rangle + i(\cos^2\theta - \sin^2 \theta) \vert 2 \rangle$.
If we consider perfect beam splitters, the final state is simplified as $\vert \psi_{fin} \rangle = - \vert 1 \rangle$. The final interference is constructive in the horizontal arm, and completely destructive in the vertical arm.

\section{\label{sec:level3} Multiport beam splitter array}

\begin{figure}[!]
\centering
 \includegraphics[width=0.6\linewidth]{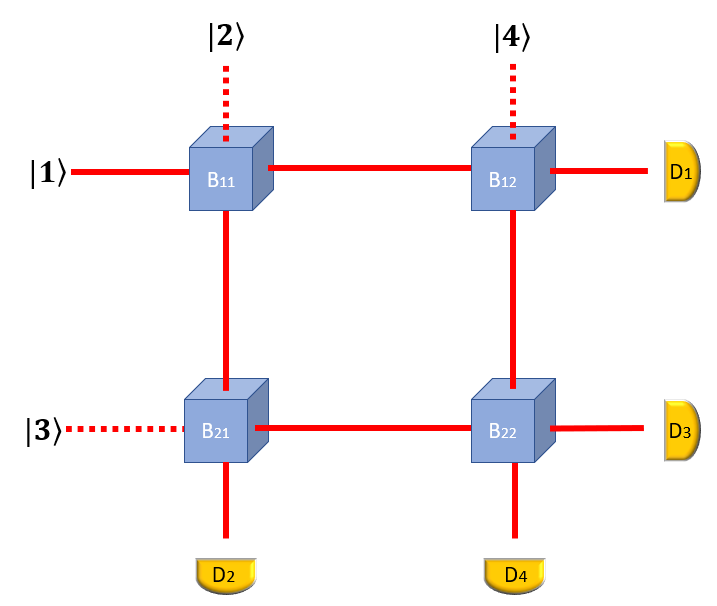}
	\caption{Beam Splitter array with four beam splitter. In this array we consider four inputs and the same number of output ports.} 
	\label{fig:bs44entradas4salidas}
\end{figure}

Inspired by this optical configuration, we propose the replacement of the two mirrors in the Mach-Zehnder interferometer by two beam splitters (See Figure \ref{fig:bs44entradas4salidas}) and the incorporation of the new exits and entrances to play a role in the photon evolution.

The name of each beam splitter in the array is in concordance with the matrix language $B_{ij}$, where $i$ stands for the row and $j$ for column position of the beam splitter in the array. In this network there are four possible entrances and an equal number of exits. As a result, the Hilbert space is expanded, and the positional states $\vert i\rangle$, where $i=1,2,3,4$, describe the potential channels through  which the photon can travel. These vectors also form a basis of the state space. The matrix representation of this basis is provided in \ref{M_4BS_array}.

 \begin{eqnarray}\label{M_4BS_array}
   \vert 1 \rangle  = 
	\begin{pmatrix}
		1 \\
		0 \\
		0 \\
		0
	\end{pmatrix},  \hskip1ex  
	\vert 2 \rangle  = 
	\begin{pmatrix}
	0 \\
	1 \\
	0 \\
	0
	\end{pmatrix},    \hskip1ex 
	\vert 3 \rangle  = 
	\begin{pmatrix}
	0 \\
	0 \\
	1 \\
	0
	\end{pmatrix},  \hskip1ex 
	   \vert 4 \rangle  =  
	\begin{pmatrix}
		0 \\
		0 \\
		0 \\
		1 \\
	\end{pmatrix}.
 \nonumber
 \\
\label{Ket1_4}
\end{eqnarray}

The action of each of these beam splitters over the possible arriving  states should be as follows: 

 \begin{eqnarray}\label{}
	&&B_{11} \vert 1 \rangle  =  \cos\theta \vert 1 \rangle  + i  \sin\theta  \vert 2 \rangle \nonumber \\
	&&B_{11}  \vert 2 \rangle =   \cos\theta \vert 2 \rangle  + i  \sin\theta  \vert 1  \rangle \nonumber \\ \nonumber \\
	&&B_{12}  \vert 1 \rangle =   \cos\theta \vert 1 \rangle  + i  \sin\theta  \vert 4  \rangle \nonumber \\
	&&B_{12}  \vert 4 \rangle =   \cos\theta \vert 4 \rangle  + i  \sin\theta  \vert 1  \rangle \nonumber \\ \nonumber \\
	&&B_{21}  \vert 2 \rangle =   \cos\theta \vert 2 \rangle  + i  \sin\theta  \vert 3  \rangle \nonumber \\
	&&B_{21}  \vert 3 \rangle =   \cos\theta \vert 3 \rangle  + i  \sin\theta  \vert 2  \rangle \nonumber \\
	 \nonumber \\
	&&B_{22}  \vert 3 \rangle =   \cos\theta \vert 3 \rangle  + i  \sin\theta  \vert 4  \rangle \nonumber \\
	&&B_{22}  \vert 4 \rangle =   \cos\theta \vert 4 \rangle  + i  \sin\theta  \vert 3  \rangle \nonumber
\end{eqnarray}  

Moreover, if the arriving state is traveling through a channel or channels that are not part of any of the paths associated with a specific beam splitter, it should act as an identity operator.
Thus, the beam splitters can be described by the following $4\times 4$ rotational matrices, 

\begin{widetext}
\begin{eqnarray}\label{M_4BS_matrices}
	B_{11}  = 
	\begin{pmatrix}
		\cos\theta & i\sin\theta & 0 & 0 \\
		i\sin\theta & \cos\theta  & 0 & 0 \\
		0 & 0 & 1 & 0 \\
		0 & 0 & 0 & 1 \\
	\end{pmatrix}, 
	B_{12}  = 
	\begin{pmatrix}
		\cos\theta & 0 & 0 & i\sin\theta \\
		0 & 1 & 0 & 0 \\
		0 & 0 & 1 &  0\\
		i\sin\theta & 0 & 0 & \cos\theta  
	\end{pmatrix}, \nonumber \\
		B_{21}  = 
	\begin{pmatrix}
		1 & 0 & 0 & 1 \\
		0 & \cos\theta & i\sin\theta & 0 \\
		0 & i\sin\theta & \cos\theta &  0\\
		1 & 0 & 0 & 1  
	\end{pmatrix},  
		B_{22}  = 
		\begin{pmatrix}
		1 & 0 & 0 & 0 \\
		0 & 1 & 0 & 0 \\
		0 & 0 & \cos\theta & i\sin\theta\\
		0 & 0 & i\sin\theta & \cos\theta 
	\end{pmatrix}.
	\nonumber
\end{eqnarray} 
\end{widetext}

For instance, let's consider the  initial state of an individual photon as $\vert \psi_{ini} \rangle  = \vert 1 \rangle $. The resulting output state of the photon after interacting with the beam splitter network can be obtain by performing the following operations: $B_{22}B_{21}B_{12}B_{11} \vert 1 \rangle$. It is important to highlight that the order in which the beam splitters are applied to the input state to obtain the output state is significant. Moreover, it is worth noting that beam splitters commute along the ascending diagonal. Finally, through further analysis, we can derive the final expression for the state (See equation \ref{FinState_BS4}).

\begin{widetext}
\begin{equation}\label{FinState_BS4}
	\vert \psi_{fin} \rangle = \cos^2\theta \vert 1 \rangle + i\sin\theta \cos\theta \vert 2 \rangle - 2 \sin^2\theta \cos\theta \vert 3 \rangle + (i \sin\theta \cos^2\theta - i\sin^3\theta) \vert 4 \rangle
\end{equation}
\end{widetext}

\begin{figure}[!ht]
	\centering	\includegraphics[width=0.7\linewidth]{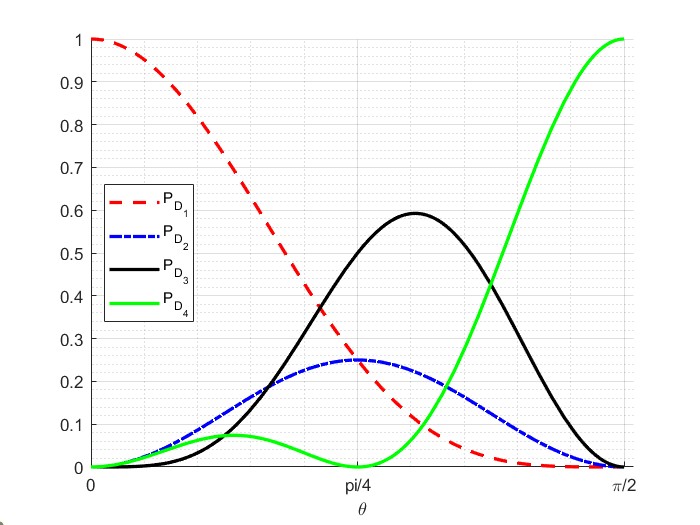}
	\caption{Probabilities in detector $D_1$ (dashed red), in detector $D_2$ (dotted Blue), in detector $D_3$ (solid black), in detector $D_4$ (solid green)}
	\label{fig:proba4bstheta}
\end{figure}

Consequently, the photon has different probabilities to be detected in each detector and they depend on the beam splitters used in the array. For example, if we consider a perfect beam splitter, i.e., $\theta=\frac{\pi}{4}$ the only detector with zero probability is $D4$; however, the other three detector have a probability different from zero (See figure \ref{fig:proba4bstheta}).  

Note that in this exemplification of the problem, the parameter $\theta$ has been set to the same value for all beam splitters. Different instantiations can be achieved by easily setting distinct values of $\theta$ for each beam splitter operator $B_{i,j}$. Therefore, it is possible to control the final state of a photon in a four-beam splitter array, as the probability of detection at each detector varies with the $\theta$ parameter. Additionally, the conventional Mach-Zehnder interferometer is a particular case of this example problem.

\section{\label{sec:level6} Generalizing the multiport beam splitter array}

\begin{figure}[!ht]
\centering
\includegraphics[width=0.6\linewidth]{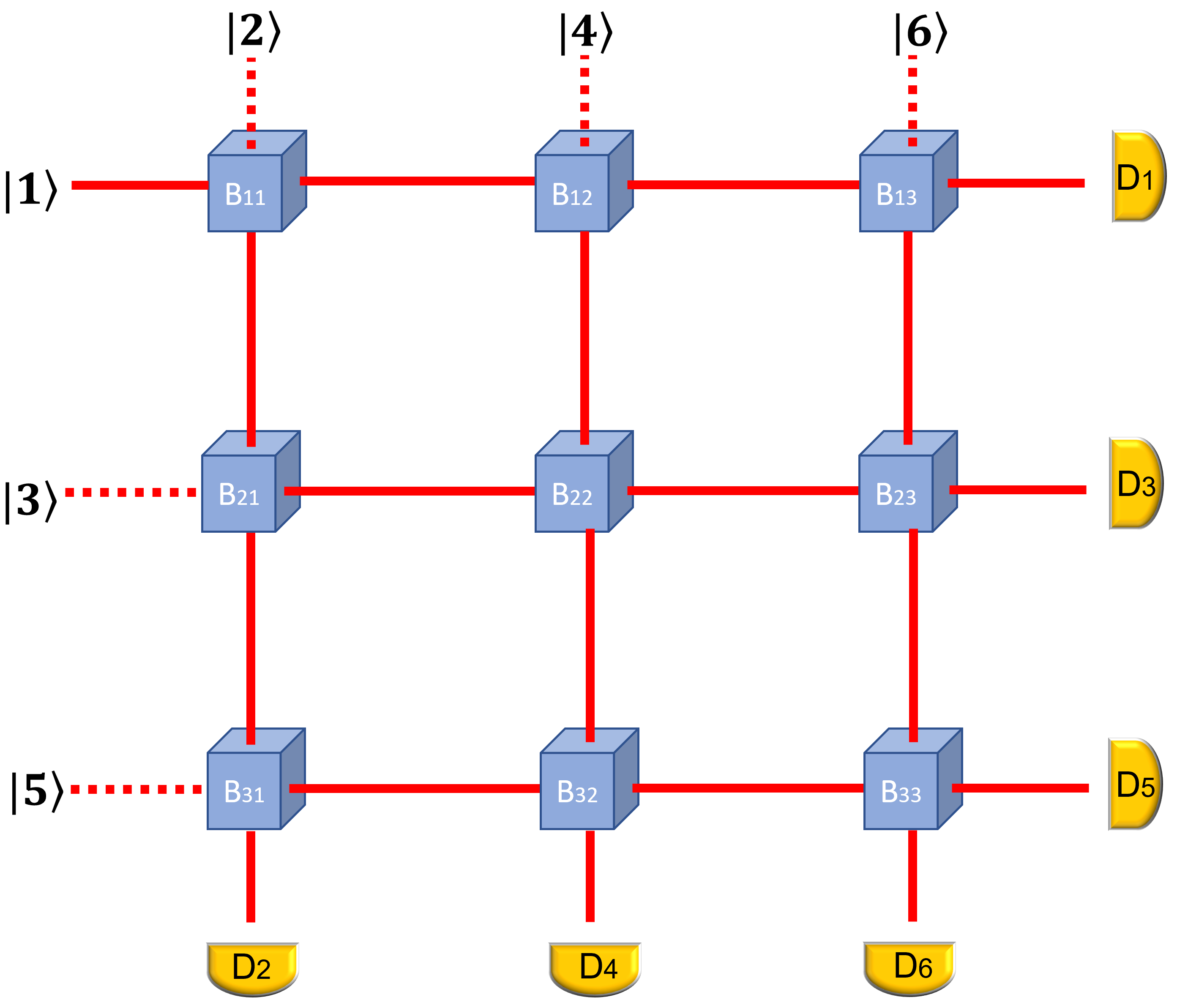}
    \caption{Depiction of a beam splitter array characterized by $p=3$. The array has $3^2$ beam splitters, the square matrices that describe each of these beam splitters are of size $6\times 6$, see equation \ref{BnmGeneral}.}
\label{fig:bs96entradas6salidas}
\end{figure}

In this section, we present a generalization of the previous array. The system is described using a parameter $p\in \mathbb{N}$, indicating that we have a square array consisting of $p^2$ beam splitters and $2p$ channels. Consequently, the Hilbert space dimension is $2p$, denoted as ${\cal{H}} = \{\vert n \rangle\vert n\in \mathbb{N}\leq 2p \}$. For this reason, the beam splitter representations belong to $2p\times 2p$ matrices. For example, refer to Figure \ref{fig:bs96entradas6salidas} where a system with $p=3$ is depicted.

The beam splitters in this larger array must adhere to a similar set of rules. They should split signals that arrive at them and act as identity operators for signals that do not. Additionally, there should be sets of beam splitter operators that commute with each other as long as they belong to the same ascending diagonal. Consequently, the operator describing the action of each beam splitter is given as follows (See equation \ref{BnmGeneral}).

\begin{widetext}
\begin{eqnarray}
		B_{m,n}= \mathbbm{1}_{2p\times 2p} + (\cos\theta -1)\left[ \vert 2n \rangle \langle 2n \vert + \vert 2m-1 \rangle \langle 2m-1 \vert\right] + i \sin\theta \left[\vert 2n \rangle \langle 2m -1 \vert  + \vert 2m-1 \rangle \langle 2n \vert\right]
		\label{BnmGeneral}
\end{eqnarray}
\end{widetext}

On the other hand, it is advantageous to introduce the general operator, $MZ$, which takes into account the sequential application of the beam splitter operators from the top-left to the bottom-right of the array. Consequently, the complete array operator is defined as follows (See equation \ref{MZoperator}).

\begin{widetext}
\begin{equation}
MZ=\left[\prod_{r=1}^{p-1} \left( \prod_{s=1}^{f(r)} B_{p-s+1,p+s-r}\right)\right] \left[\prod_{r=p}^{2p-1} \left( \prod_{s=1}^{f(r)} B_{p-s+1,s}\right)\right]
\label{MZoperator}
\end{equation}
\end{widetext}
	
\noindent
where $f(r)$ is,

\begin{equation}
	f(r) = \left\{ \begin{array}{ll}
		r & \mbox{if $r \leq p$};\\
		2p-r & \mbox{if $r > p$}.\end{array} \right.
		\label{fr}
\end{equation}

\noindent
By offering the flexibility to adjust transmission/reflection properties through manipulation of the corresponding $\theta$ values, the operator described in equation \ref{MZoperator} becomes a powerful tool. We can indicate which of these devices are identity operators, mirrors, beam splitters or even, with certain configuration, complete absorbing objects. Furthermore, it is possible to analyze both, square and non-square arrays, by setting some of these devices as identity operators. 

\section{\label{sec:level4} Instances of our model}

In this section, we explore specific instances that exemplify the versatility and robustness of our model. Firstly, we will examine the simplest examples where the output is known based on reported experimental and theoretical realizations, see for example \cite{Eli93,Kwi95}. Then, we will proceed to more intricate systems that, to the best of our knowledge, have not been experimentally realized yet.

As a first example, we consider the output of a Mach-Zehnder interferometer (See Figure \ref{figMZ}) when an incoming photon enters the system through port $1$. In other words, the initial state is $\vert 1 \rangle$.

\begin{figure}[!ht]
\centering
\includegraphics[width=0.6\linewidth]{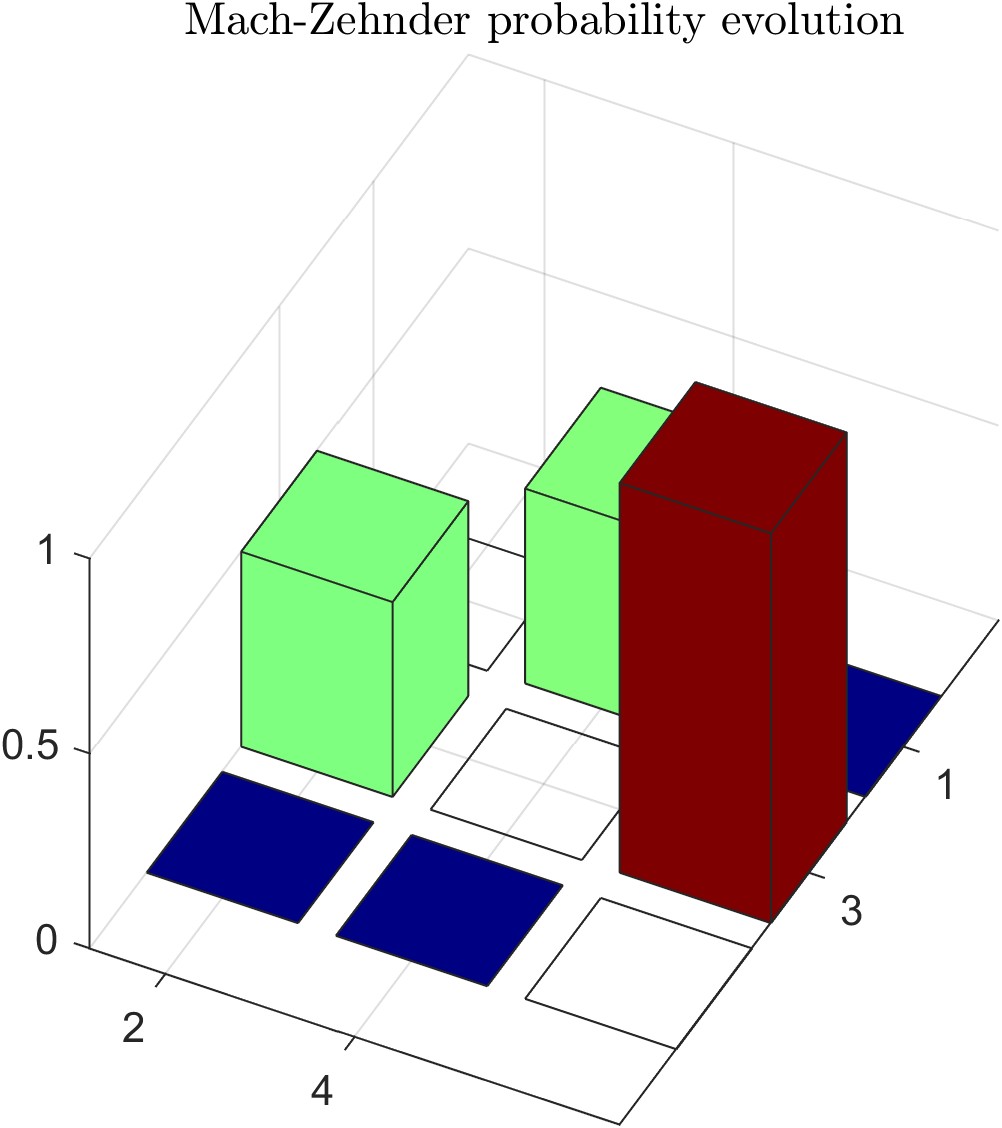}
    \caption{Probability evolution of a photon on a Mach-Zehnder interferometer using the operator $MZ$ (equation \ref{MZoperator}) with an initial state $\vert \psi_{ini} \rangle = \vert 1 \rangle$. The final probabilities of the $1,3$ and $2,4$ channels are shown in the right column and lower row, respectively. The label $3$ corresponds to detector $D_1$, and the label $4$ corresponds to detector $D_2$. Refer to figure \ref{figMZ} for further details.}
\label{Ejemplo1}
\end{figure}

\noindent
By applying the operator defined in equation \ref{MZoperator}, considering $\theta=\frac{\pi}{4}$, to the initial state $\vert \psi_{ini} \rangle = \vert 1 \rangle$, we can obtain the same probabilities that are reported in a Mach-Zehnder array, section \ref{sec:level2}. As shown, in Figure \ref{Ejemplo1}, when the state interacts with the first beam splitter, the photon has an equal probability of $0.5$  to reach each of the mirrors. Since the mirrors are fully reflective, there is a probability of $0$ for the signal to continue to $ch 2$ or $ch1$. Therefore, as the state reaches the next beam splitter, it exhibits total destructive interference to continue along $ch 4$ and total constructive interference to $ch3$. Consequently, the photon has a probability of $0$ to be detected in $ch4$ (corresponding to detector $D_2$ according to Figure \ref{figMZ}), and a probability of $1$ to be detected in $ch 3$ (corresponding to detector $D_1$). This information is summarized in Figure \ref{Ejemplo1}, which presents the complete probability evolution in the Mach-Zehnder array. 

\subsection{\label{subsec:level1} Rectangular beam splitter arrays}

\begin{figure}[!ht]
\centering
\includegraphics[width=0.6\linewidth]{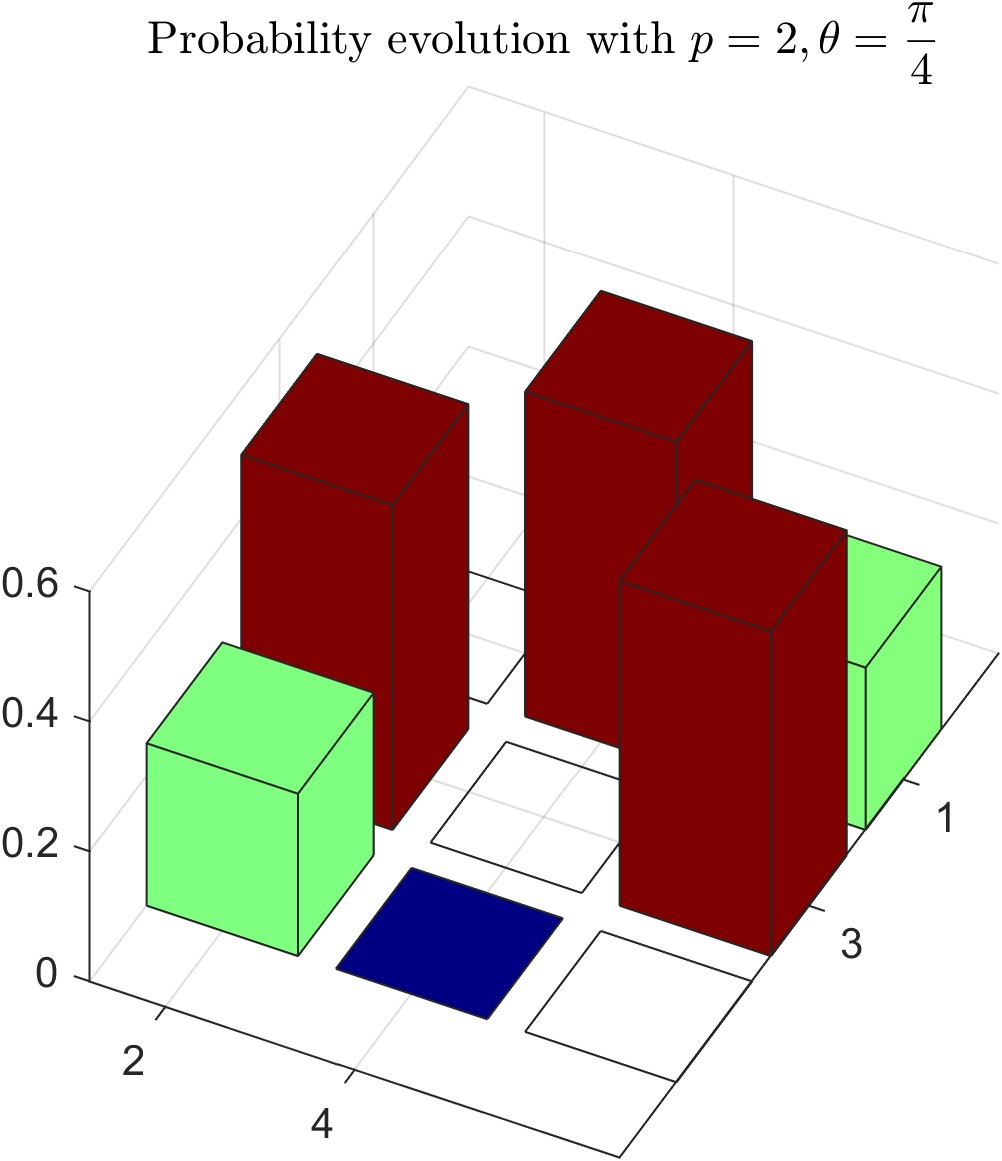}
    \caption{Evolution probability of one photon on a beam splitter array, $p=2$, $\theta=\pi/4$, $\vert \psi_{ini} \rangle = \vert 1 \rangle$.}
\label{Ejemplo2}
\end{figure}

When the two perfect mirrors in the Mach-Zehnder interferometer, each corresponding to $\theta=\pi/2$, are replaced by perfect beam splitters with $\theta=\pi/4$ (see Figure \ref{fig:bs44entradas4salidas}), the probability evolution undergoes a change. The complete evolution of the probability can be seen in Figure \ref{Ejemplo2}.

\noindent
Once again, the probabilities at each detector can be observed at the end of their respective paths. In Figure \ref{Ejemplo2}, the green rectangle on the left represents the probability at detector 2, as indicated in Figure \ref{fig:bs44entradas4salidas}. Similarly, the blue rectangle corresponds to the probability at $D_4$, the red rectangle represents $D_3$, and finally, the green rectangle on the right represents $D_1$. The first two red rectangles, from top to bottom, indicate the probabilities after interacting with the first beam splitter but before reaching the subsequent components. Figure \ref{Ejemplo2} provides a comprehensive view of the probability evolution in the beam splitter array.

\begin{figure}[!ht]
\centering
\includegraphics[width=0.6\linewidth]{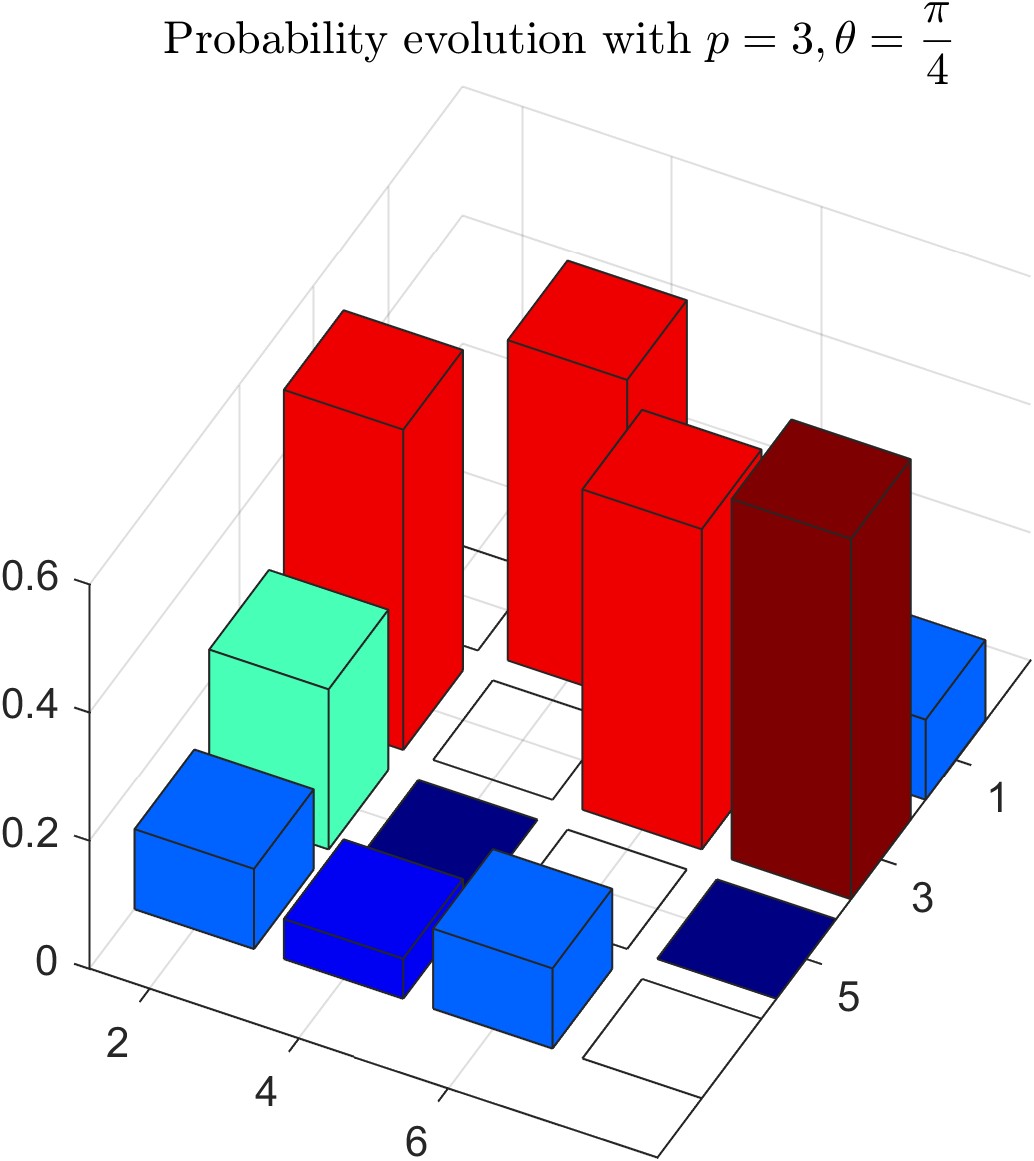}
    \caption{Probability evolution  of a single photon on a beam splitter array. In this case $p=3$ and each beam splitter has been set to $\theta=\pi/4$. The initial state is $\vert \psi_{ini} \rangle = \vert 1 \rangle$.}
\label{Ejemplo3}
\end{figure}

Furthermore, we can analyze the evolution of a single photon in larger arrays by exploring different values of the parameter $p$. For instance, let's consider the case where $p=3$, which corresponds to a system with 9 beam splitters. In this scenario, we will also set the value of $\theta$ to $\pi/4$ for each of the beam splitters. The complete probability evolution is presented in Figure \ref{Ejemplo3}.

In Figure \ref{Ejemplo3}, we observe the evolution of the system with $p=3$ and a similar set of perfect beam splitters. The response begins with two red squares in the top left, representing the first beam splitter interaction. Subsequently, we see a sequence of green, blue, and red rectangles, followed by another green rectangle consistent with the result obtained for $p=2$. Finally, the front row and right column display the probabilities of photon detection at detectors $D_2$, $D_4$, $D_6$, and $D_5$, $D_3$, $D_1$, respectively.

\begin{figure}[!ht]
\centering
\includegraphics[width=0.6\linewidth]{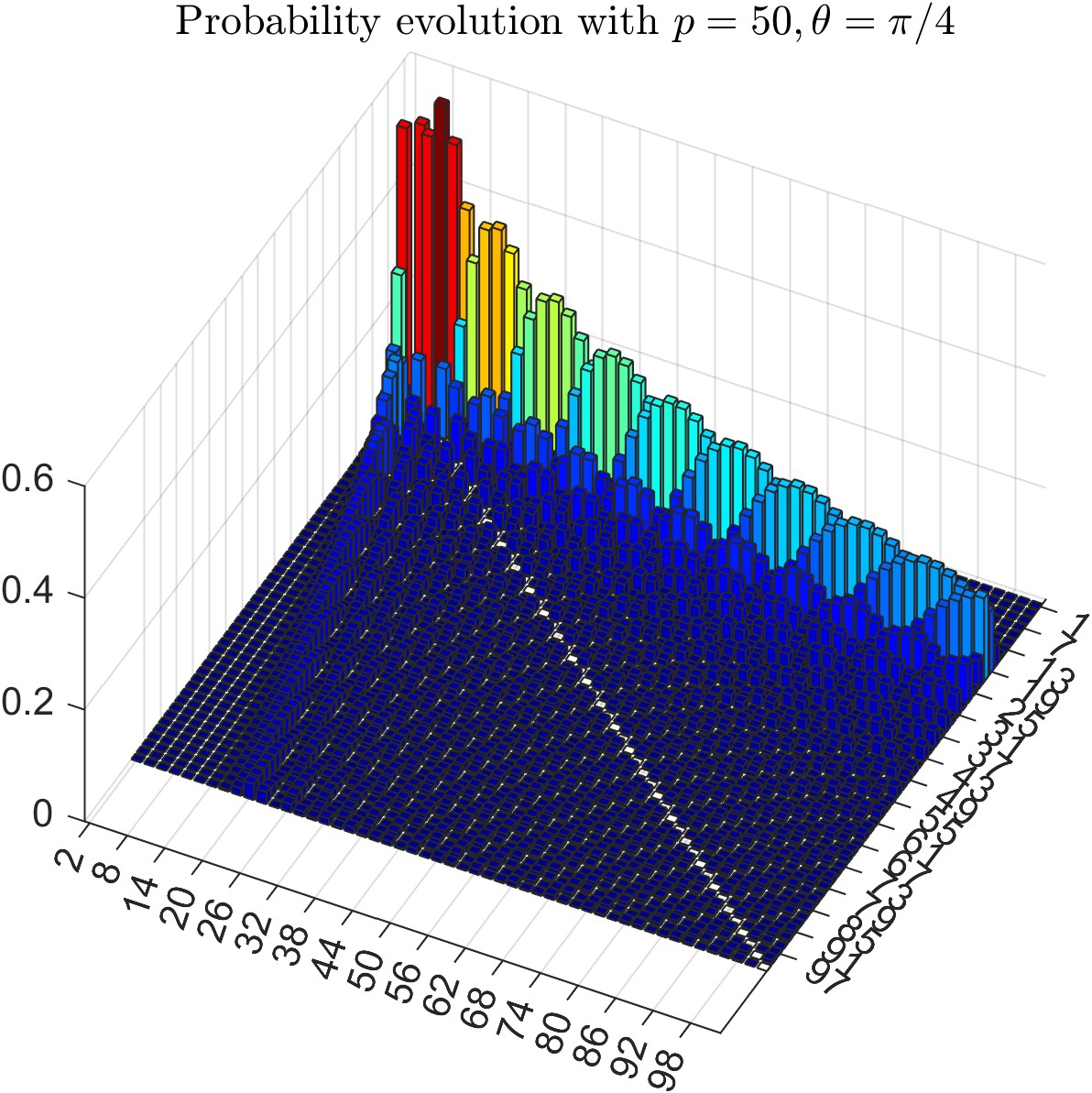}
    \caption{Probability evolution of a single photon on a beam splitter array with initial state $\vert \psi_{ini} \rangle = \vert 1 \rangle$.}
\label{Ejemplo4}
\end{figure}

In a similar vein to the previous example, we investigated the system's evolution with $p=50$, which corresponds to an array of $2500$ beam splitters, each set to $\theta=\pi/4$. The probability evolution is depicted in Figure \ref{Ejemplo4}.

\begin{figure}[!ht]
\centering
\includegraphics[width=0.6\linewidth]{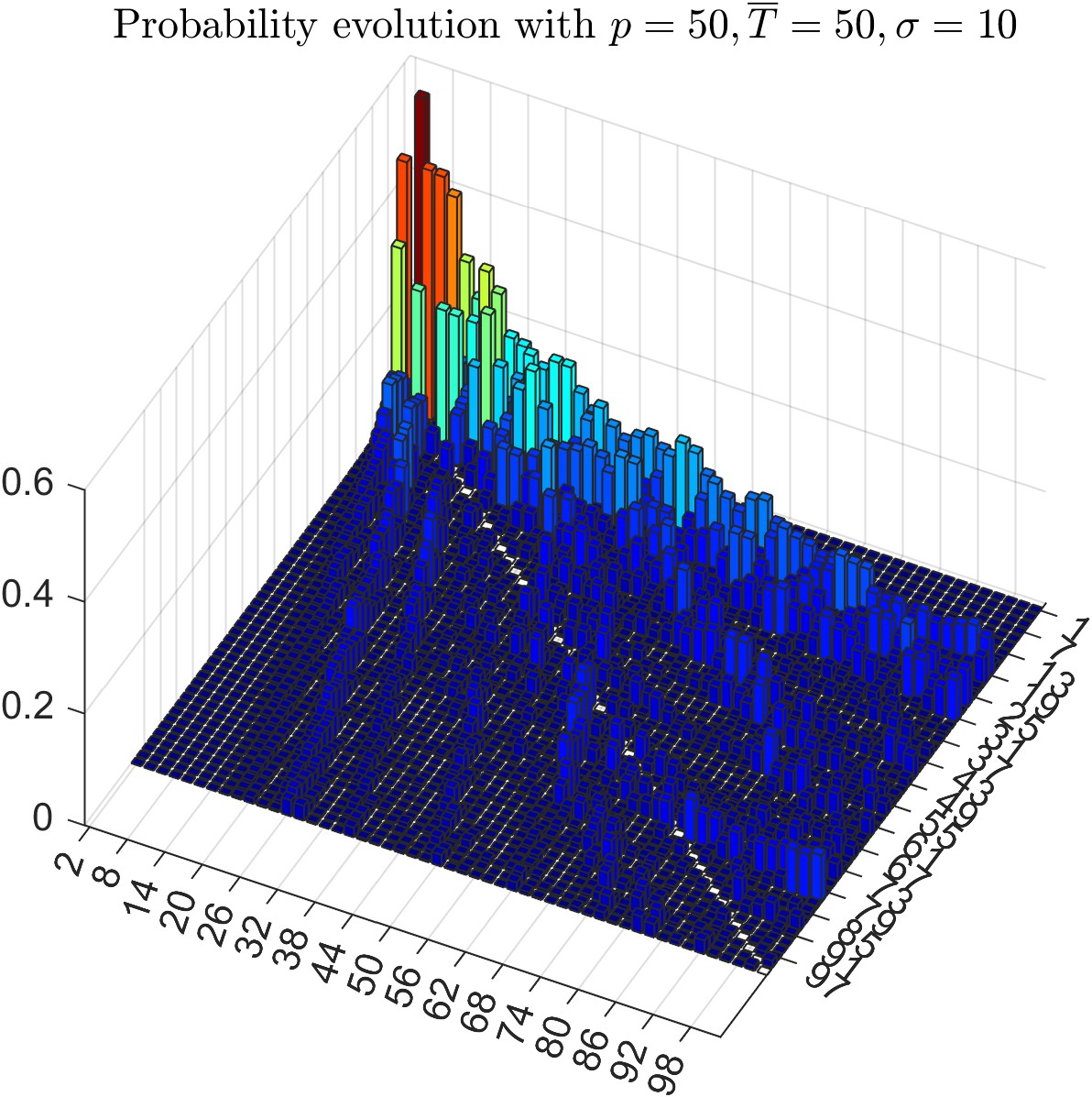}
    \caption{Probability evolution of a single photon on a beam splitter array with average transmission coefficient $\overline{T} = 50$ (with $\sigma = 10$) and initial state $\vert \psi_{ini} \rangle = \vert 1 \rangle$.}
\label{Ejemplo5}
\end{figure}

Perfect beam splitters do not exist in real life, and the transmission coefficients of individual beam splitters can vary due to various factors such as imperfections in manufacturing, temperature variations, or aging effects. To account for this variability, we introduce randomness in the transmission coefficients of each beam splitter in our model. Specifically, we generate the transmission coefficient for each beam splitter using a normal distribution centered at 50 with a standard deviation of 10. By incorporating this randomness, we can observe the system's response under more realistic conditions. The corresponding results, considering the variability in the transmission coefficients, are shown in Figure \ref{Ejemplo5}.

It is evident from Figure \ref{Ejemplo5} that the probability evolution follows an unordered path, in contrast to the pattern observed in Figure \ref{Ejemplo4}.

\begin{figure}[!ht]
\centering
\includegraphics[width=0.6\linewidth]{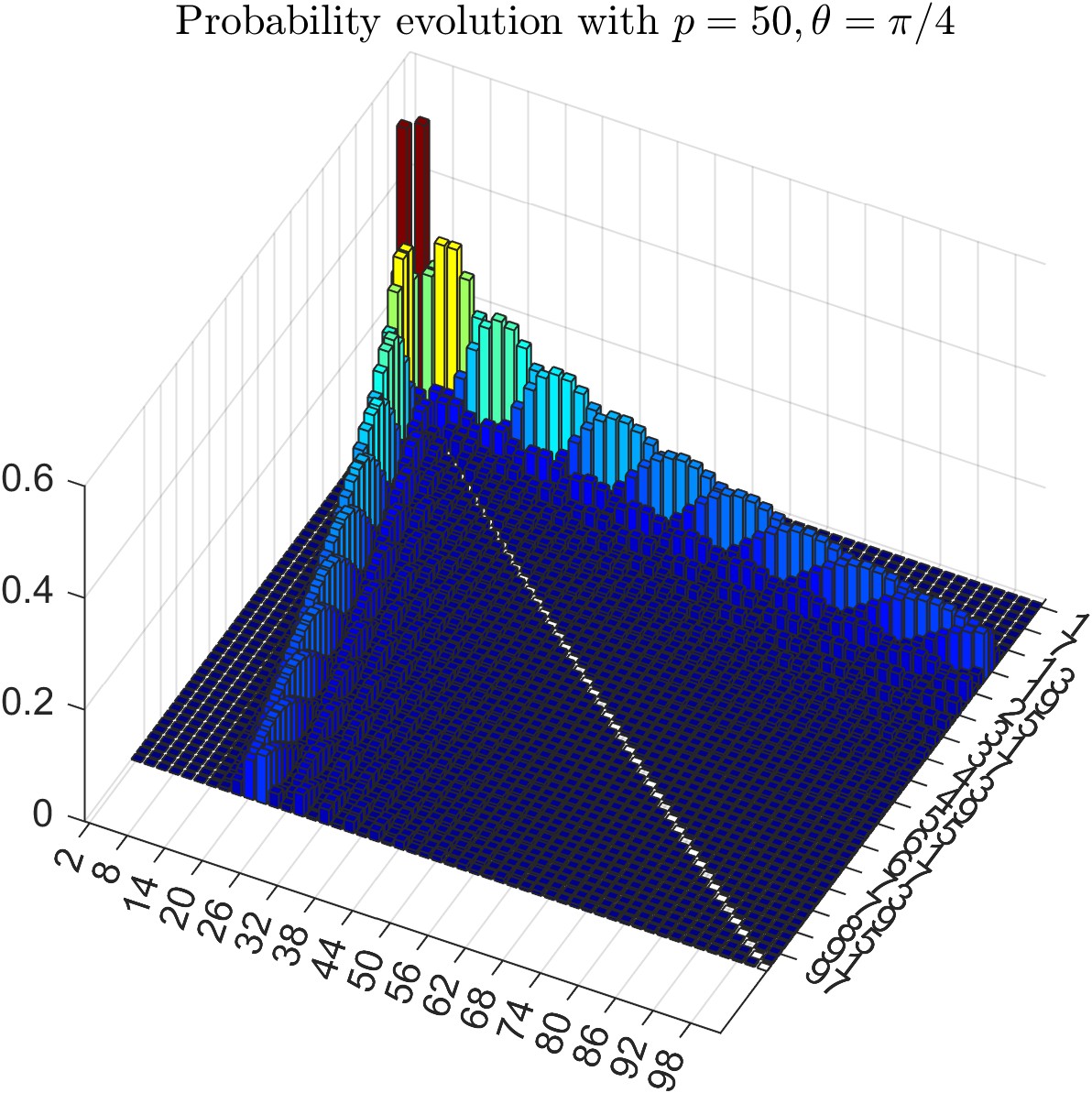}
    \caption{Probability evolution of a single photon in a beam splitter array. The initial state is $\vert \psi_{ini} \rangle = \frac{1}{\sqrt{2}}\left(\vert 1 \rangle+\vert 2 \rangle\right)$. }
\label{Ejemplo6}
\end{figure}

Based on the results depicted in Figure \ref{Ejemplo5}, the output probabilities deviate from a balanced distribution despite the well-balanced transmission and reflection coefficients of each beam splitter with $\theta=\pi/4$. The photon demonstrates a clear tendency to continue its path towards the odd-numbered detectors. This observation leads us to explore, as our penultimate rectangular example, the behavior when the input state is a linear combination of $\vert 1\rangle$ and $\vert 2\rangle$. The system's response to this scenario is illustrated in Figure \ref{Ejemplo6}.

\begin{figure}[!ht]
\centering
\includegraphics[width=0.6\linewidth]{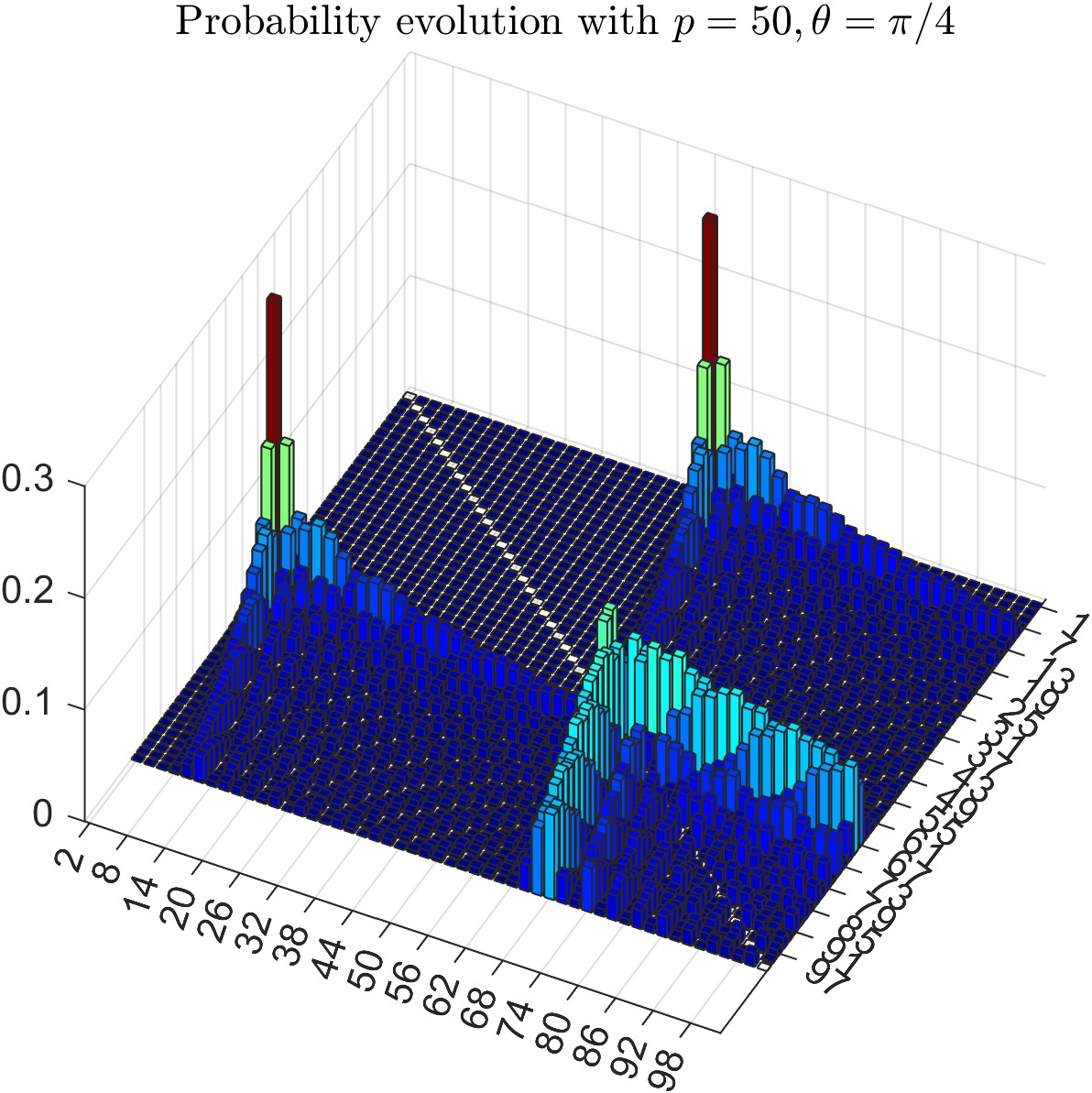}
    \caption{Probability evolution of a single photon in a beam splitter array. The initial state is $\vert \psi_{ini} \rangle = \frac{1}{\sqrt{2}}\left(\vert 49 \rangle+\vert 50 \rangle\right)$. }
\label{Ejemplo7}
\end{figure}

As our last rectangular example, we explore our system response when the input state is a linear combination of $\vert 49\rangle$ and $\vert 50\rangle$, see Figure \ref{Ejemplo7}.

In addition to these specific instances, the general beam splitter array model presented in Section \ref{sec:level6} allows for the exploration of a wide range of configurations and scenarios, even non square configurations that can be achieve by simple setting $\theta=0$ and avoiding such beam splitters to perform operations in the incoming states so that it arrives directly to the detectors. By varying the parameters $p$ and $\theta$ and analyzing the probability evolution, one can investigate the behavior of photons in different network architectures and study their potential applications in quantum information processing, communication, and other fields.

\subsection{\label{subsec:level1} Non-square beam splitter arrays}

\begin{figure}[!ht]
\centering
\includegraphics[width=0.5\linewidth]{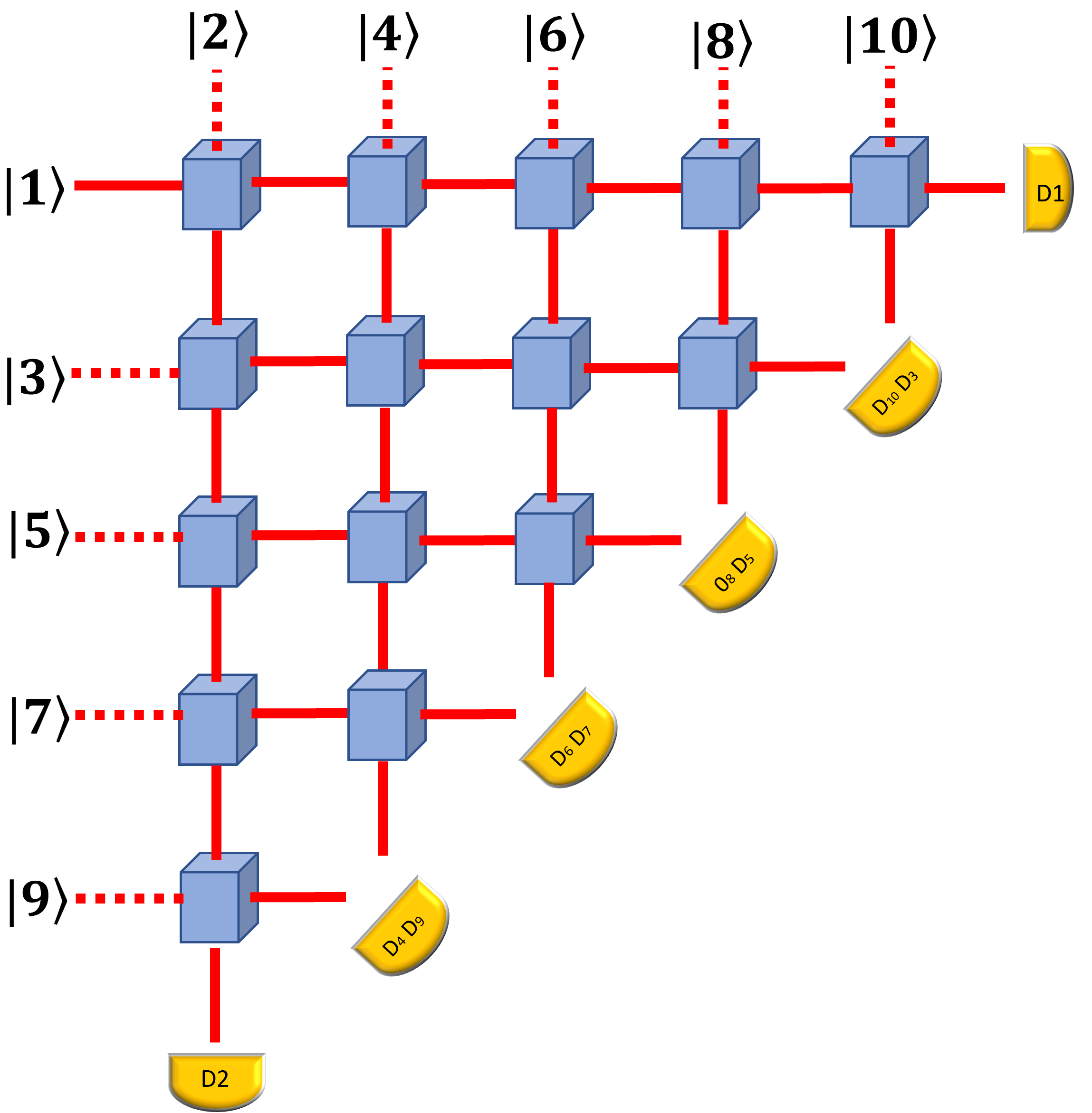}
    \caption{Galton board: non-square array of beam splitters. The beam splitters above the diagonal act as identity operators.}
\label{GB}
\end{figure}

An illustrative instance of a non-square beam splitter array is the Galton board, as depicted in figure \ref{GB}. Within this array the beam splitters positioned below the diagonal are defined as identity operators.  

\begin{figure}[!ht]
\centering
\includegraphics[width=0.45\linewidth]{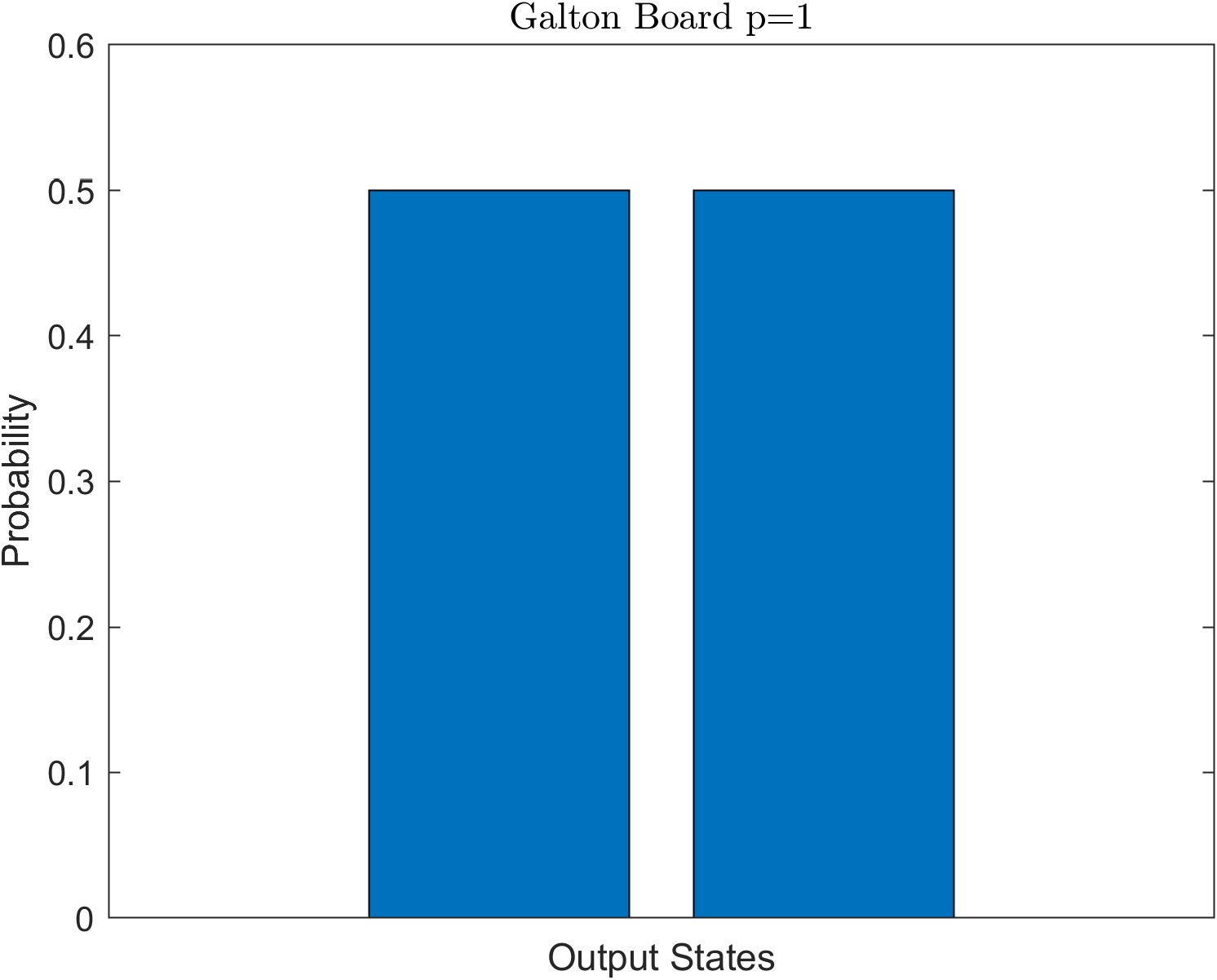}
\includegraphics[width=0.45\linewidth]{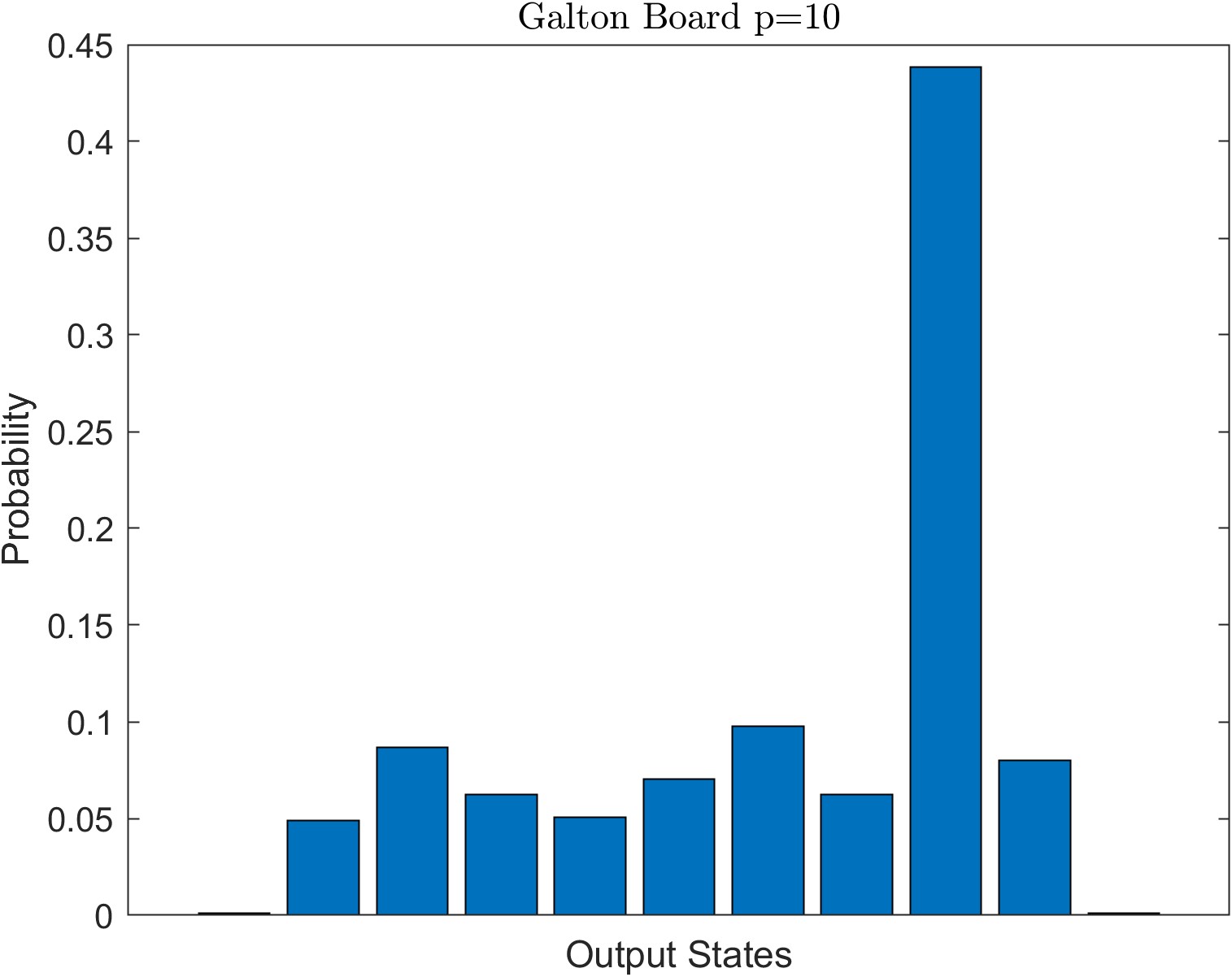}
\includegraphics[width=0.45\linewidth]{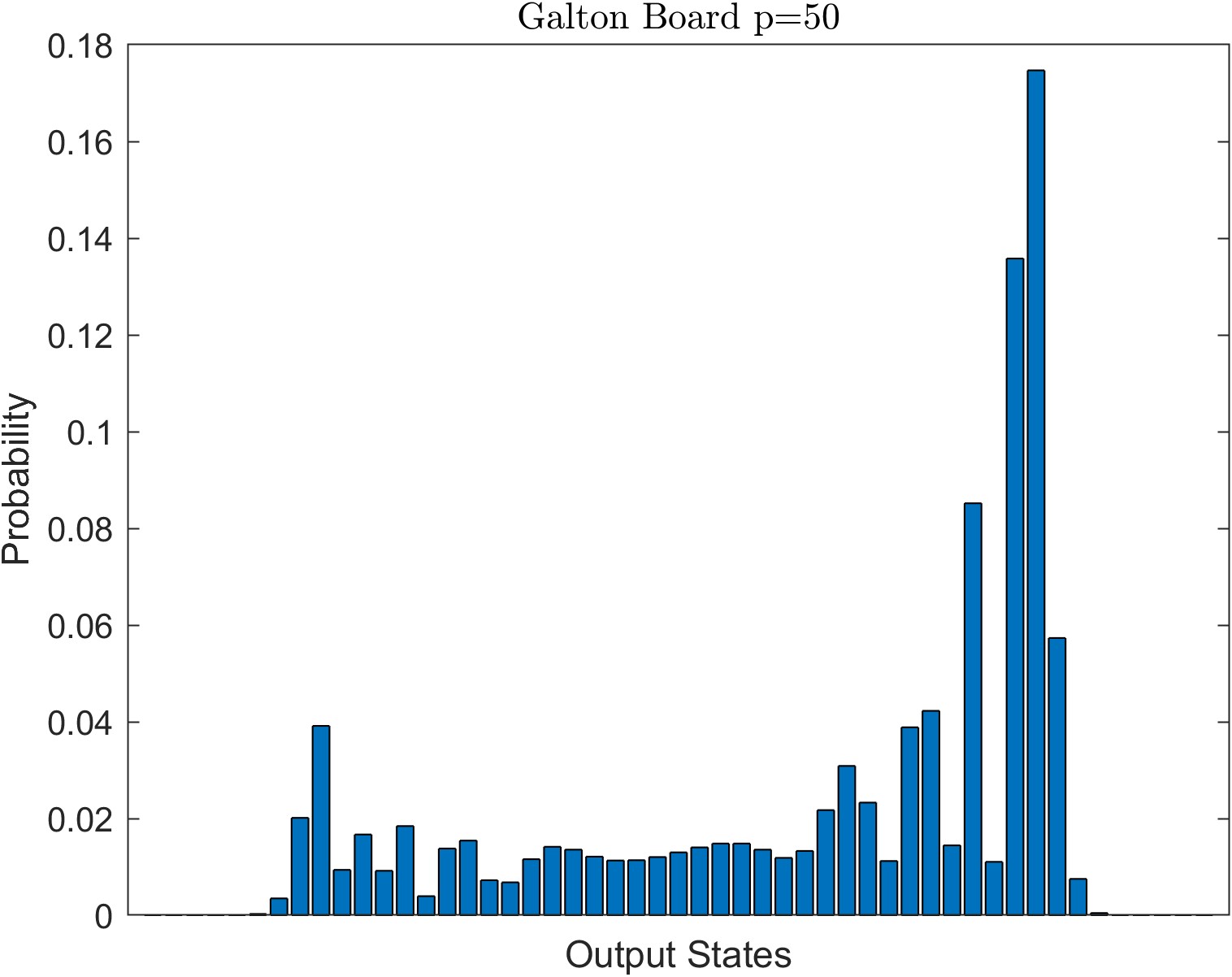}
\includegraphics[width=0.45\linewidth]{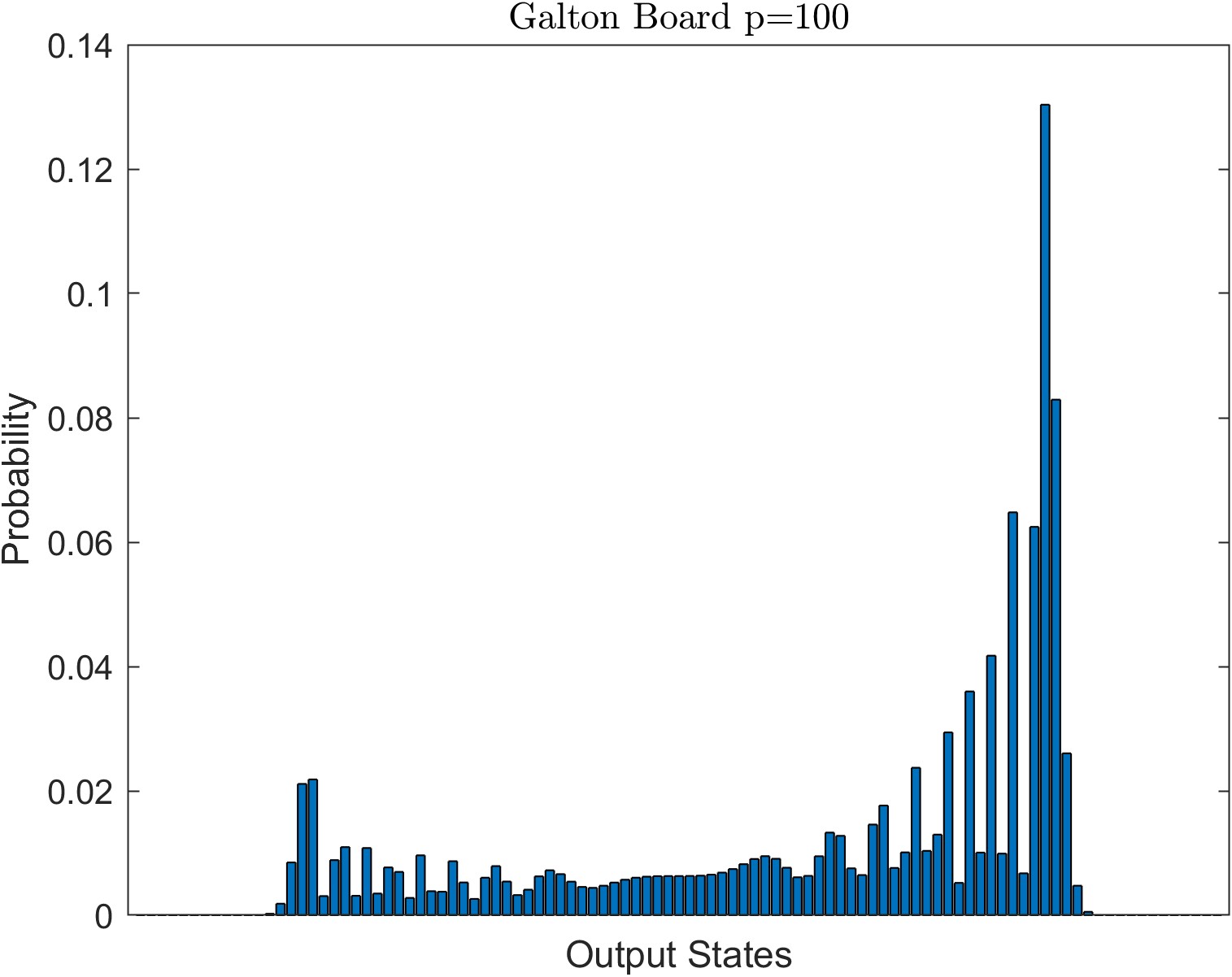}
    \caption{Probability  distributions with initial state $\vert \psi_{ini} \rangle =\vert 1 \rangle$.}
\label{PGB}
\end{figure}

In Figure \ref{GB}, the detectors have different position compared to the rectangular array. For example, the second detector (from top to bottom) receives information from channels $3$ and $10$. The position of these detectors serves the purpose of comparing the results with the classical outcome. The classical Galton board follows a binomial distribution; however, when an individual photon travels on it, the behaviour changes, and it can be accurately described using our general model. In Figure \ref{PGB}, we account for the Galton board's responses of various sizes.

It is important to note that if the initial state is  $\vert \psi_{ini} \rangle =\vert 2 \rangle$, a significant change occurs in the probability distribution. The most probable outcomes shift from the upper detectors to the lower ones, resulting in a mirror-like pattern of the graphs shown in Figure \ref{PGB}.

\begin{figure}[!ht]
\centering
\includegraphics[width=0.45\linewidth]{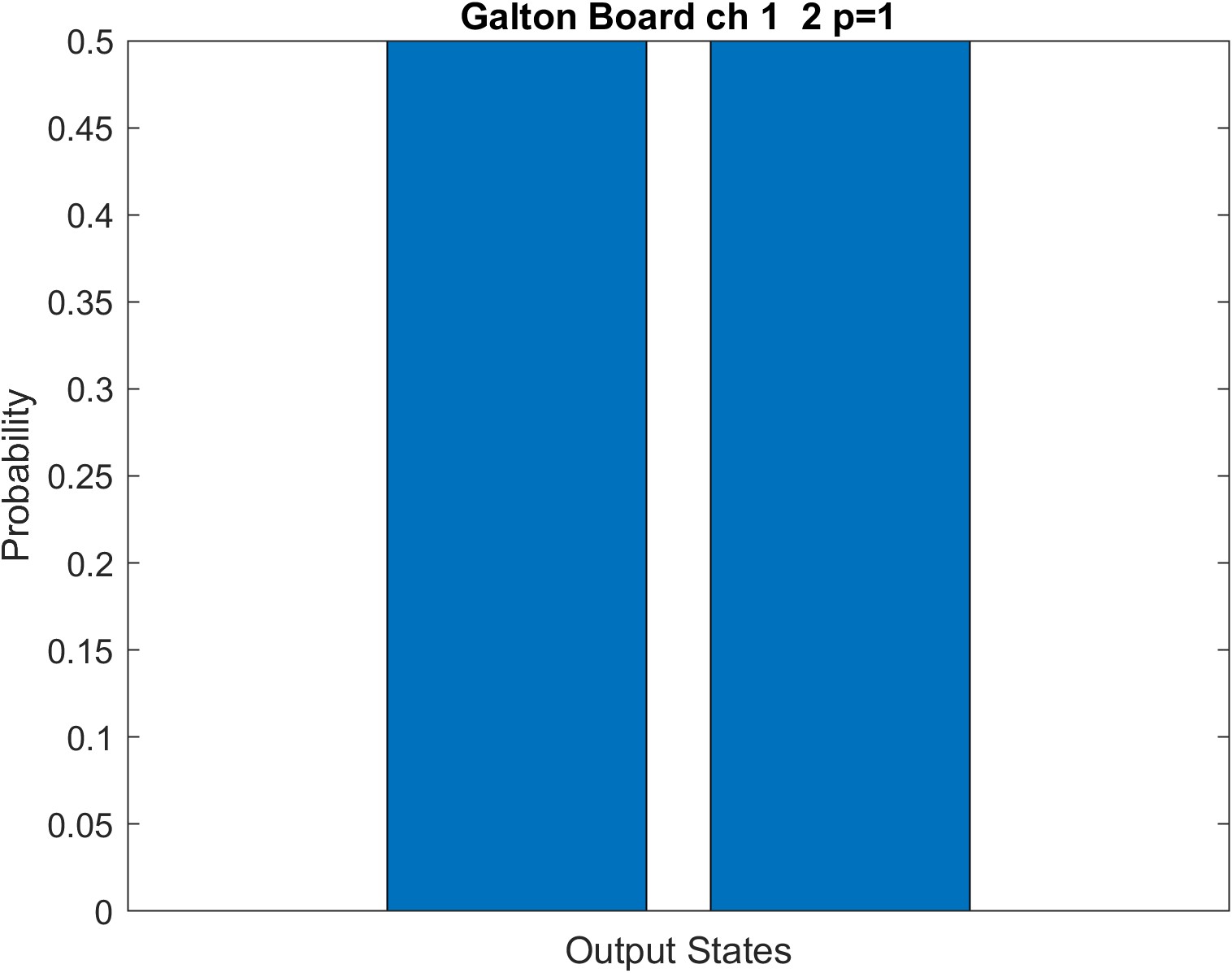}
\includegraphics[width=0.45\linewidth]{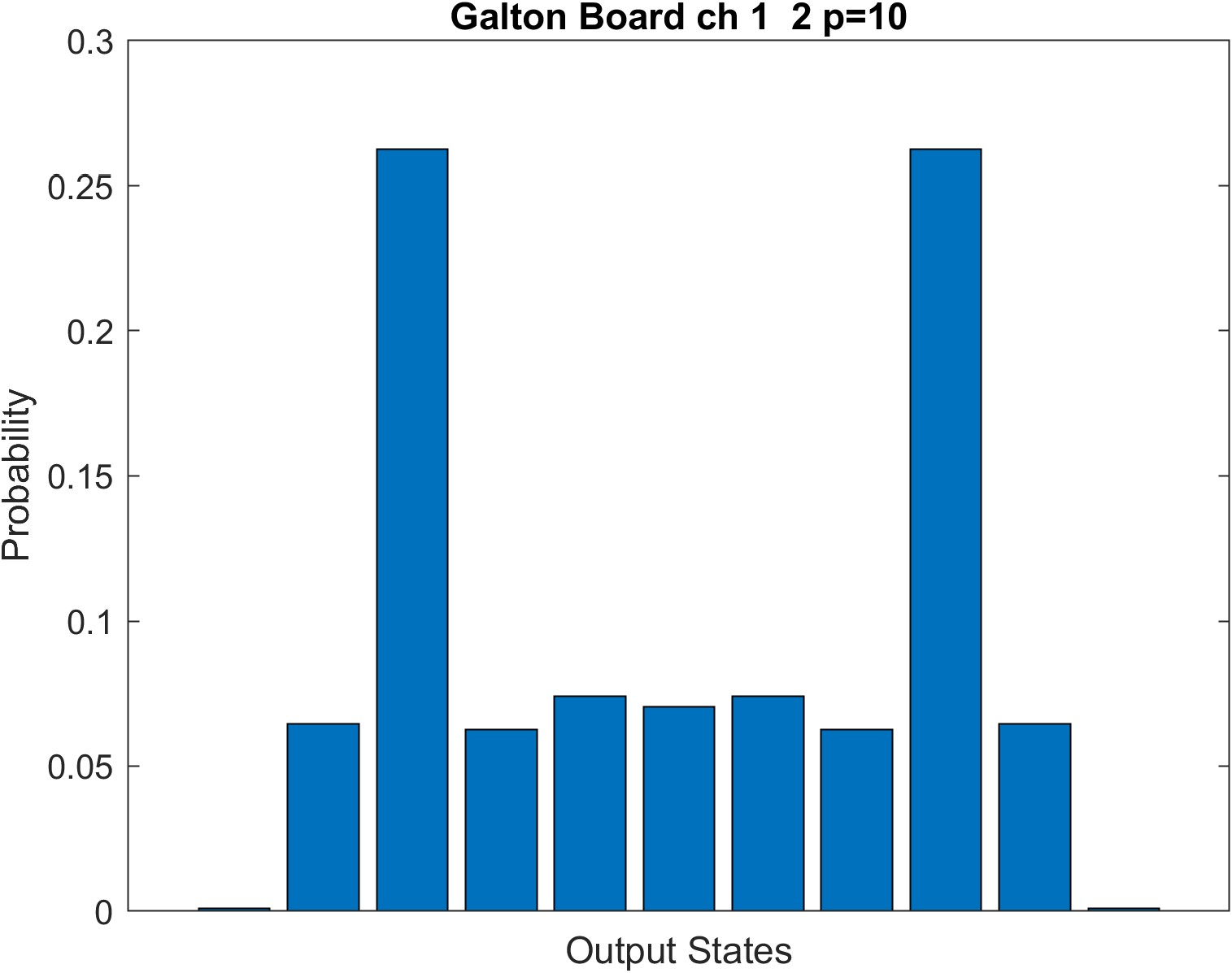}
\includegraphics[width=0.45\linewidth]{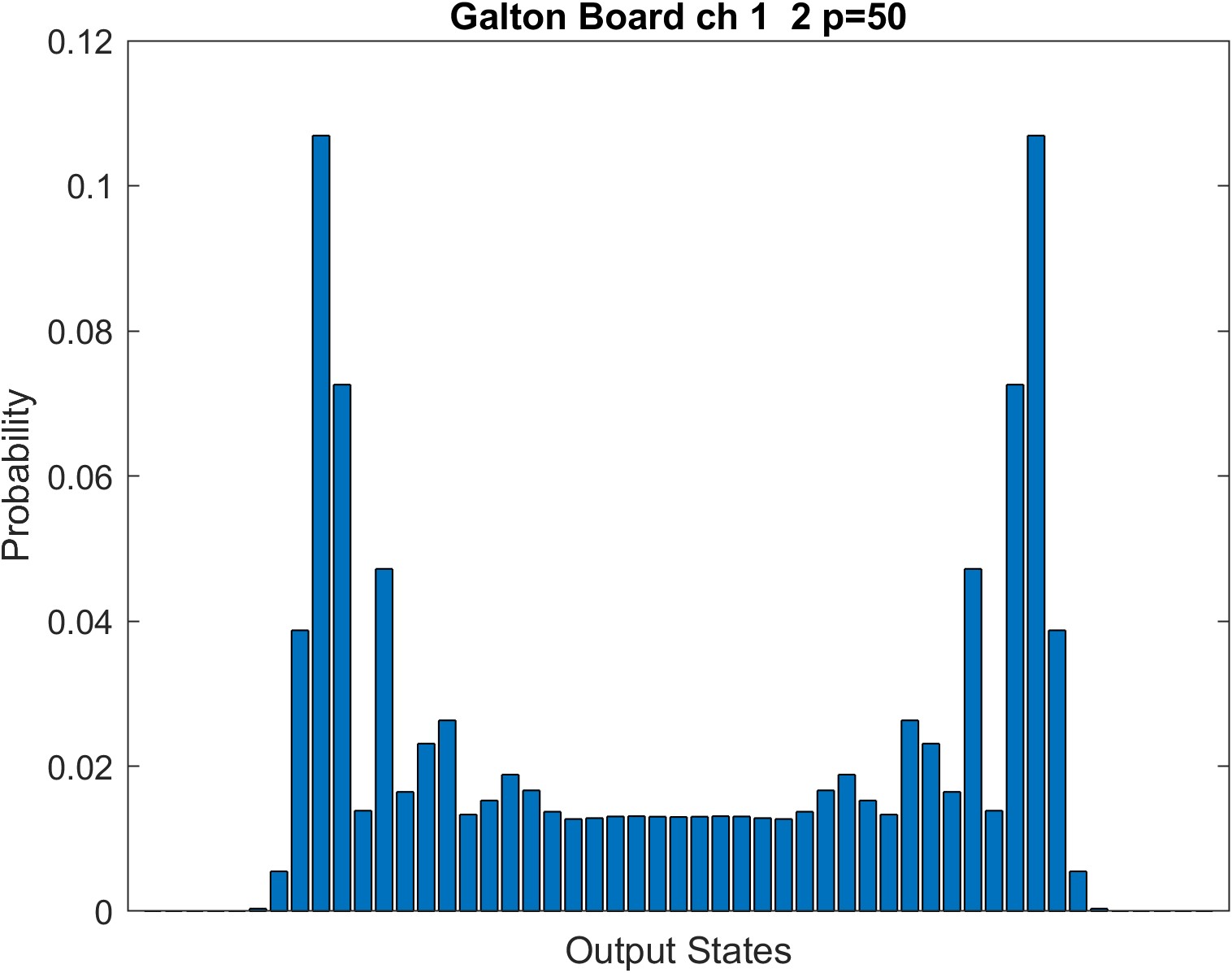}
\includegraphics[width=0.45\linewidth]{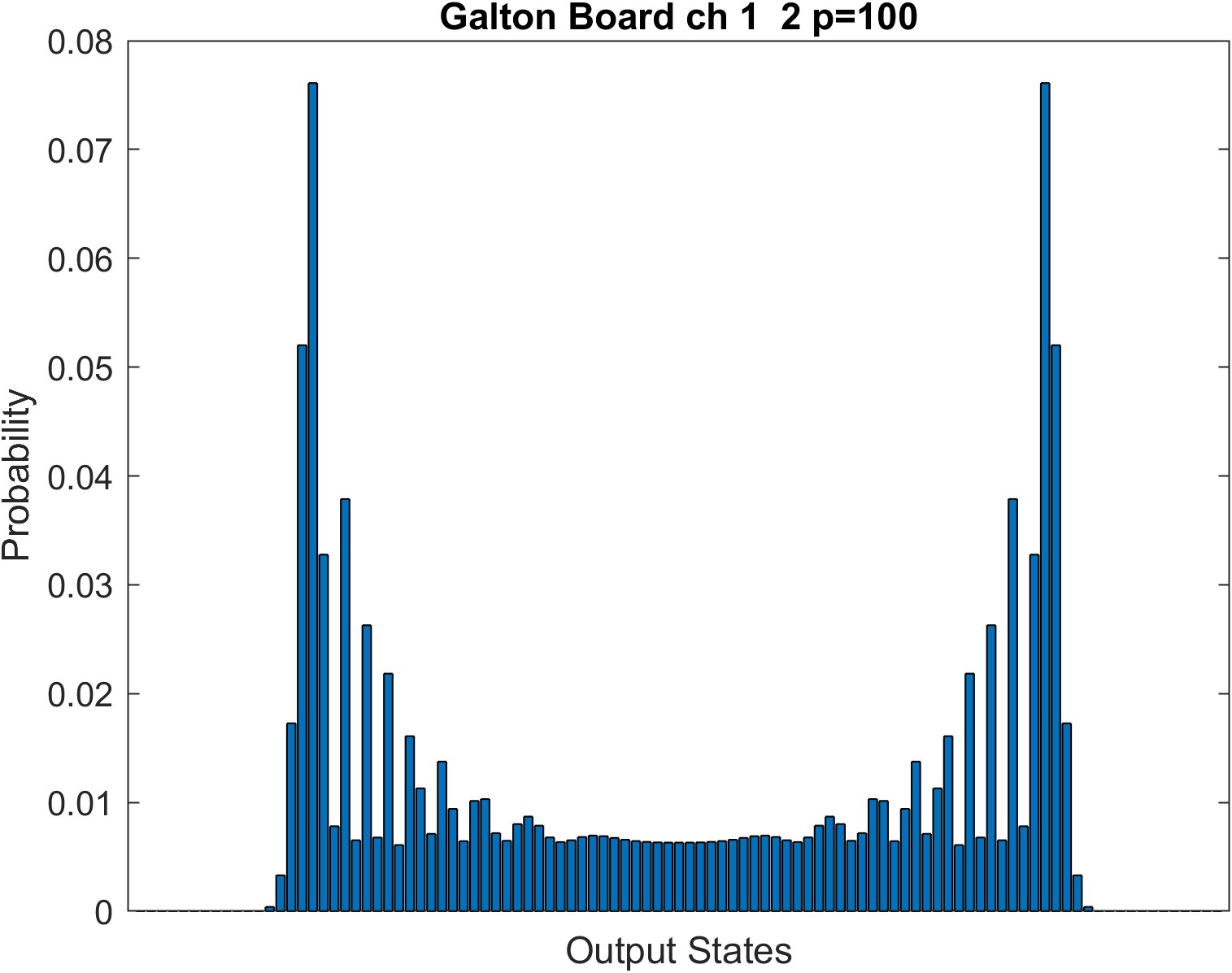}
    \caption{Probability  distributions with initial state $\vert \psi_{ini} \rangle = \frac{1}{\sqrt{2}}\left(\vert 1 \rangle+\vert 2 \rangle\right)$}.
\label{PGB_LC}
\end{figure}

Following a similar scheme as before, we can prepare an initial state as a superposition of states: $\vert \psi_{ini} \rangle = \frac{1}{\sqrt{2}}\left(\vert 1 \rangle+\vert 2 \rangle\right)$. The resulting probability distribution is depicted in Figure \ref{PGB_LC}. 

\begin{figure}[!ht]
\centering
\includegraphics[width=0.9\linewidth]{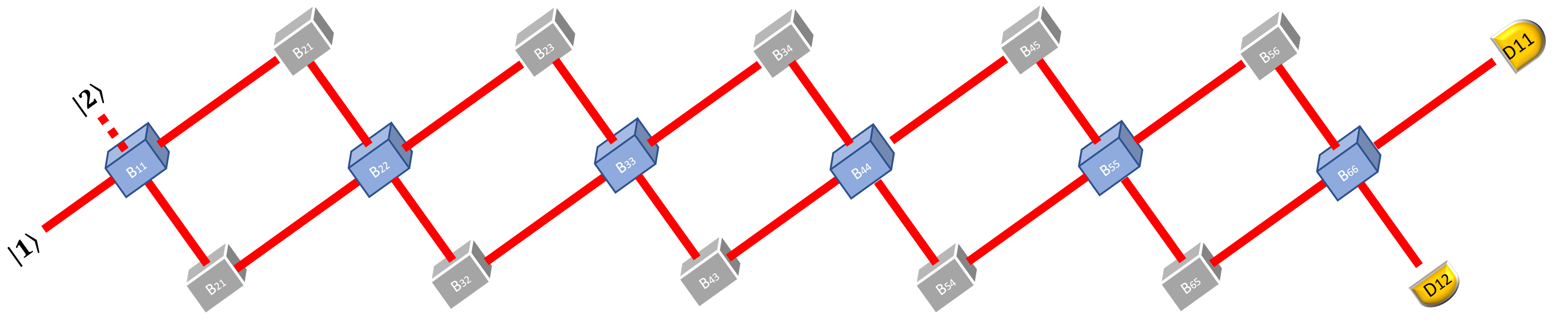}
\includegraphics[width=0.9\linewidth]{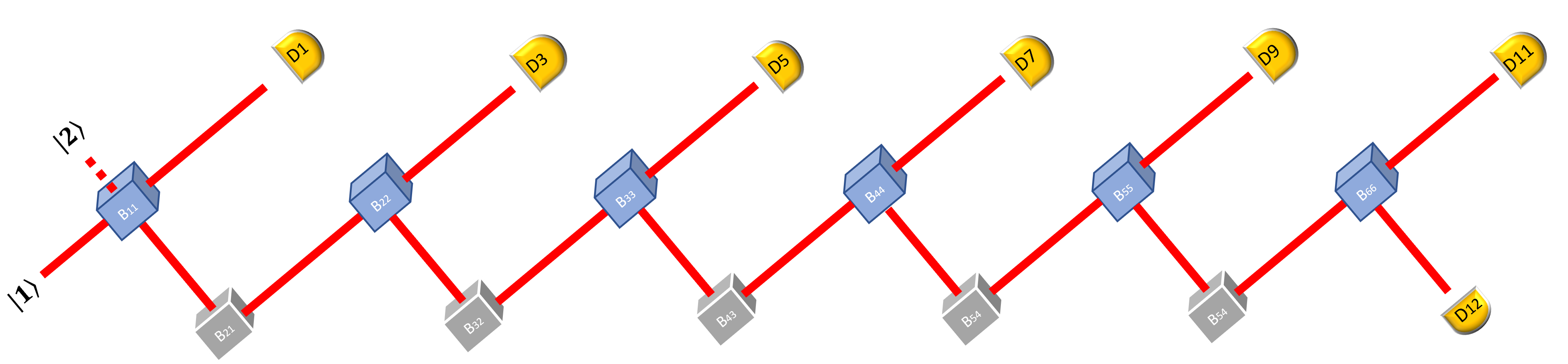}
    \caption{Set up for the discrete quantum Zeno effect. Upper and lower diagrams showcase scenarios where the absence or presence of an object could interrupt the interference  \cite{Kwi95}. Mirrors are depicted by the gray boxes, where $\theta = \frac{\pi}{2}$. Blue boxes are beam splitters with $\theta=\frac{\pi}{2}-\frac{\pi}{2p}$; in this case $p=6$.}
\label{QZE}
\end{figure}

As our final example, we will replicate the system presented in \cite{Kwi95} to illustrate the discrete Quantum Zeno effect (see Figure \ref{QZE}). This apparatus involves a non-rectangular array of mirrors and beam splitters, with transmission and reflection coefficients adjusted to $\theta = \frac{\pi}{2}-\frac{\pi}{2p}$. In our framework this non-square array must be incorporated into a square configuration. The additional beam splitters, not depicted in this diagram, are set to $\theta = 0$, representing identity operators.

\begin{figure}[!ht]
\centering
\includegraphics[width=0.45\linewidth]{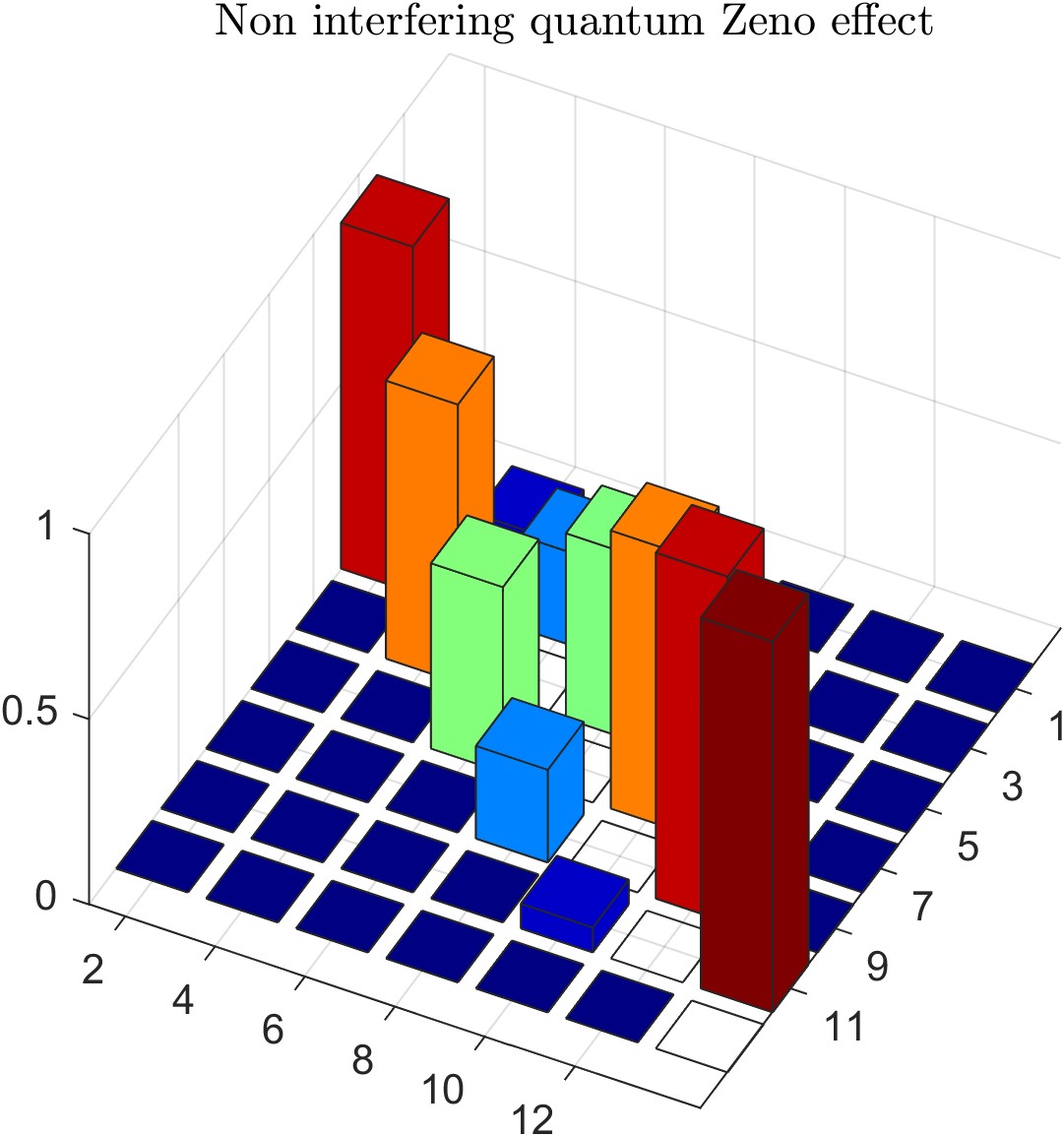}
\includegraphics[width=0.45\linewidth]{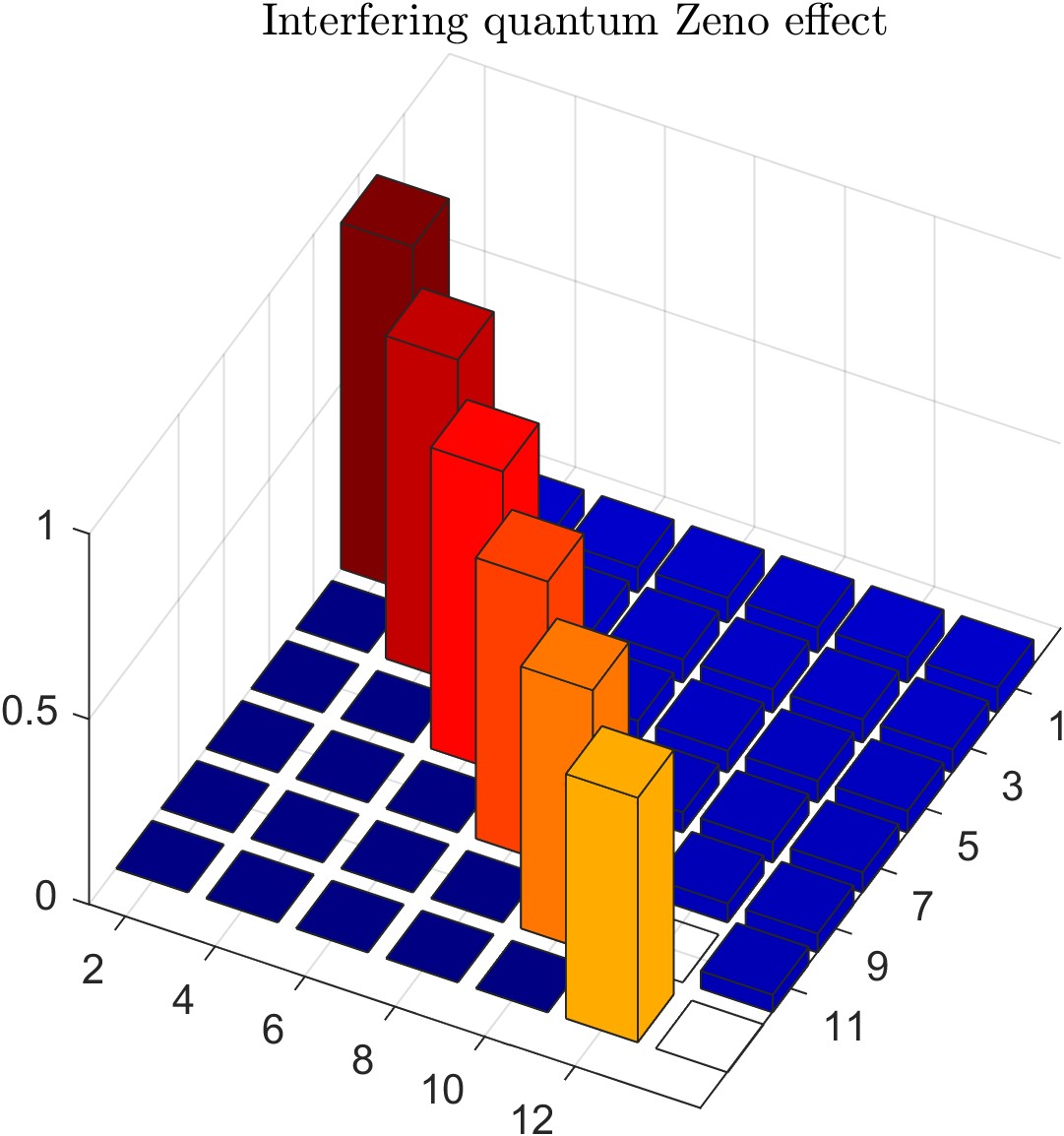}
    \caption{Quantum Zeno effect evolution without and with repeated measurement, respectively.}
\label{QZER}
\end{figure}

As a result, we can observe the effects of the absence or presence of the object in the form of a repeated measurement in upper paths, see Figure \ref{QZER}.

As discussed in \cite{Kwi95}, the probability of photon detection shifted significantly from detector $11$ to detector $12$, such probability is no longer zero. The remaining probability, now distributed in the odd-numbered detectors, corresponds to the repeated measurements or, more precisely, the likelihood of interaction with the object. In a real-world experiment, the activation of detector $12$ with a single photon would exemplify an interaction-free measurement, a concept pioneered by Elitzur and Vaidman \cite{Eli93}.

\section{\label{sec:level5} Conclusion}

We have introduced an advanced matrix representation for beam splitter arrays.  The inherent beam splitters are characterized by rotation matrices, each precisely tailored to operate on individual channels. In our study, we have showcased various example problems, ranging from fundamental setups like the Mach-Zehnder interferometer to more intricate systems such as the quantum quincunx and the discrete quantum Zeno effect.

Future research will focus on exploring different architectures and the incorporation of Fock states into the system.

\nocite{*}

\bibliography{apssamp}

\end{document}